\documentclass[10pt]{article}

\usepackage{amsmath}
\usepackage{amssymb}

\usepackage{graphicx}

\usepackage{cite}

\usepackage{color}
\usepackage {url}

\usepackage[labelfont=bf,labelsep=period,justification=raggedright]{caption}

\makeatletter
\renewcommand{\@biblabel}[1]{\quad#1.}
\makeatother

\date{}

\begin{document}

\begin{flushleft}
{\Large
\textbf{Finite Element Procedures for Enzyme, Chemical Reaction and '\textit{In-Silico}' Genome Scale Networks}
}
\\
{\small Martins, R.C.$^{1}$, Fachada, N.$^{2}$}
\\
\vspace{0.25cm}
$^1$ \small Life and Health Sciences Research Institute (ICVS) and \\
\small\ ICVS/3B’s - PT Government Associate Laboratory, Braga/Guimarães, Portugal \\
\small Universidade do Minho, Campus of Gualtar, 4710-057 Braga-Portugal
\\
$^2$ \small ISR – Institute for Systems and Robotics, Instituto Superior Técnico,\\
\small Av. Rovisco Pais, 1, 1049-001 Lisboa, Portugal \\
$\ast$ E-mail: rui.martins@ecsaude.uminho.pt
\end{flushleft}

\section*{Abstract}

The capacity to predict and control bioprocesses is perhaps one of the most important objectives of biotechnology.
Computational simulation is an established methodology for the design and optimization of bioprocesses,
where the finite elements method (FEM) is at the state-of-art engineering multi-physics simulation system,
with tools such as Finite Element Analysis (FEA) and Computational Fluid Dynamics (CFD).

Although FEA and CFD are currently applied to bioreactor design, most simulations are restricted to
the multi-physics capabilities of the existing sofware packages. This manuscript is a contribution for
the consolidation of FEM in computational biotechnology, by presenting a comprehensive
review of finite element procedures of the most common enzymatic mechanisms found in biotechnological
processes, such as, enzyme activation, Michaelis Menten, competitive inhibition, non-competitive inhibition,
anti-competitive inhibition, competition by substrate, sequential random mechanism, ping-pong bi-bi
and Theorel-Chance.

Most importantly, the manuscript opens the possibility for the use of FEM in conjunction with «in-silico» models of
metabolic networks, as well as, chemical networks in order to simulate complex bioprocesses in biotechnology,
putting emphasis into flux balance analysis, pheno-metabolomics space exploration in time and space,
overcoming the limitations of assuming chemostat conditions in systems biology computations.




\vspace{0.5cm}
\noindent \small{\textbf{Keywords}: Finite element analysis, enzyme kinetics, 'in-silico', genome scale networks}

\clearpage
\newpage
\section*{Introduction}

Predicting the behavior of bioprocesses is one of the major goals of biotechnology. Computational
simulation is today a valuated tool for predicting, monitoring and controlling the status of fermentations,
as well as, for optimizing fermentation conditions, minimizing trial and error experimental procedures.

Computational design is recognized as a standard prototyping tool outside the bioengineering area
(e.g. automotive and aviation), where it significant reduces costs during design, prototyping
and testing phases. All of these, generally involve high experimental load and trained personnel in different areas
of research and engineering. The same is also becomming a reality in biotechnology with the advent of systems
and synthetic biology.

Traditional experimental methods are limited by the number of recorded parameters for a holistic systems
characterization. The conjunction of high-throughput methods (e.g. mass spectroscopy, microarrays, sequencing,
spectroscopy and electrochemistry) are today elected for validation of state-of-the-art 'in-silico' chemical and
genome scale models (GSM). Computational simulation provides detailed information in time and space.
The Finite Element Method (FEM) is of the "heart" of many Finite Element Analysis (FEA) and Computational
Fluid Dynamics (CFD) software for simulating physical phenomena, such as, heat transfer, mass transfer, radiation,
fluid dynamics, structural and elasticity, but it can also be used in biotechnology for simulation of
chemical, biochemical reactions, and cellular dynamics \cite{Martins:2006,Martinsetal:2009b}.

FEM was not initially developed for computational biology and bioprocesses simulation. It has been
devoted to industrial prototyping of biotech machinery \cite{Davidsonetal:2003,Ghadgeetal:2006,O-Charoenetal:2007}.
It is not yet usual the application of FEM for the simulation of complex biological or chemical systems
\cite{Gelleretal:2006,Kashidetal:2007,Munthe:2000,Shepel:2006}. In sophisticated developments, FEM has been
used to compute microscopic properties, such as: i) the study of membrane elasticity
 \cite{Hansenetal:1997}: ii) electrostatic interactions between proteins \cite{Zhou:1993}; 
iii) mechanical modeling of ion channels \cite{Tangetal:2006}; iv) applying FEM in microscopy for physical
properties estimation \cite{CharrasandHorton:2002}.  
FEM has also been applied to the study of enzyme kinetics by continuous diffusional biomolecular systems given
by the Smoluchwski equation. It has proven to be a good alternative to the traditional spherical criterion model,
by allowing to study the complex enzyme geometries
\cite{Elcocketal:1996,Elcocketal:2001,GadzoulineandWade:1998,Taraetal:1998,Radicetal:1997}.

The continuity of FE facilitates the inclusion of other phenomena such as, fluid flow, heat/mass transfer,
electromagnetic field, forces and elasticity. The computational cost is less when computing large scale problems
described by differential equations, where continuous solutions are common in physical phenomena even at small
scales (e.g. force fields, diffusion, heat transfer)
\cite{DruryandDembo:1999,Sachs:1999,Songetal:2004,Watanabeetal:2006}.

The main steps in FEA involve: i) Pre-processing; ii) Resolving the PDEs or ODEs in the physical-time domain;
and iii) post-processing. Pre-processing generally involves: i) ensure that PDEs and ODEs are interactive for
multi-physics and chemical, biochemical and microbiological models; ii) ensure that the solution is stable and
accurate in the physical-time domain by optimizing the mesh refinement and time steps from computer assisted design
software \cite{Fluent:2004,Ansys:2004,CFX:2004} or in more complex geometrics (e.g. biological tissues) by digital scanning and 2D/3D
reconstruction methods. This methodology has a number of advantages, such as the treatment of
problems on complex irregular shapes, non-uniform meshing to reflect different levels of multi-scale detail,
treatment of boundary conditions using continuous solutions and the construction of higher-order approximations
to improve accuracy of numerical solutions. Both biological materials, as well as, bioreators display
irregular geometries and non-homogeneous physical-chemical properties, which makes difficult to sustain
a chemostat hypothesis. FEM not only overcomes such hurdle, but when used in conjunction with inverse problems
makes possible to minimizing the error between simulation and experimental datasets obtained in discrete
positions of space, to improve model predictions \cite{Chen:2005,Thomee:2006,Nakasoneetal:2006}.
As biological processes implie multi-physics and multi-scale simulations, it becomes essential to:
i) develop the correct relationship between physical-chemical, biochemical and microbiological models;
ii) ensure that all used model parameters are correctly determined against experimental data by inverse
methods and statistical analysis \cite{Martinsetal:2008}.

High-throughput molecular biology and analytical chemistry technologies are exponentially increasing chemical and
biological 'omics' information databases (e.g. genomics, metabolomics, transcriptomics, 
proteomics and protein interactions) (see Figure \ref{figure:IntegrativeComplexSystems}). The available information
allowed the emergence of the annotation of gene, protein and metabolic functions, as well as, regulatory mechanisms,
so that, network reconstructions of complex biological systems are today feasible. Network models gave rise to the
development of 'in-silico' network organisms, reconstructed from curing the information present in both databases
and publications \cite{Duarteetal:2004,Herrgardetal:2006}, allowing the analysis of network properties and topology, as well as, the comprehensive
analysis of cellular functions by systems biology approaches \cite{NielsenandJewett:2007}.

Connecting all mathematical models on a multi-scale and multi-physics strategy is one of the most important
challenges for understanding the complexity of chemical and biological systems \cite{Martinsetal:2008}.
This manuscript is a contribution for the basis of the use of the finite element method procedures for the
integration of enzyme kinetics, chemical and genome scale network models ('in-silico' strains) as a complex
systems multi-scale and multi-physics computational modeling research area. This communication is not a
comprehensive presentation of the finite elements method, and therefore background on numerical modeling
is necessary to make use of the presented equations. 

\section*{Materials and Methods}

\subsection*{The finite elements method}

The FEM is considerably different from the most common discretization methodologies, such as Finite Differences
(FD), Finite Volumes (FV) and Lattice-Boltzmann (LB) methods. Although elements are geometrically equal, FEM ensures
that the solution is continuous inside each element, solved by a weak solutions to a variational optimization of
a quadratic problem, being the solution inside a physical given by a piecewise continuity - the shape function.

The following steps resume the FEM methodology: i) Passing from global to local coordinates for the shape function
of such element; ii) Variational analysis - determining the solution to the variational problem by weakening the solution
inside the finite element; iii) matrix assembly of all equations; and iv) solving
\cite{StrangandFix:1973,Segerlind:1984,HenwoodandBonet:1996,Braess:1997,Moaveni:1999,Hutton:2004,Chen:2005,Thomee:2006} 
and rendering results into graphical mode \cite{Moaveni:1999,Nakasoneetal:2006,VTK:2007}.

\subsubsection*{The variational method}

Changing a Partial Differential Equation (PDE) or an Ordinary Differential Equation (ODE) into the variational
form is the main procedure for any FEM discretization. The simplest form of a variational ($V(x)$),
states for two continuous functions $h(x)$ and $v(x)$:

\begin{equation}  \label{equation:variational}
        V(x) = \int_a^b h(x)v(x)dx = 0
\end{equation}

which means that $v(x)$, a \textit{weighting} or \textit{testing} function, can be chosen to force the residuals
$h(x)$ to be zero inside the finite element interval $\left[ a,b \right]$. The variational problem is posed in the
finite element space ($\Omega$). The variational can be solved by the direct substitution of the residuals function
($h(x)$) and weighting function ($v(x)$) and minimization (Garlekin's method) or by the minimization of a linear
functional (functional method) \cite{Segerlind:1984,Rozanov:1998,Ghanem:1991,Nicolaietal:2000}.
The variational method states that there is a solution to the problem of eq. \ref{equation:variational} given by:

\begin{equation}
\mathcal{B}(u,v) = \mathcal{L}(v)
\end{equation} \\

which satisfies the solution $u=u^*$, for any trial function $v$ (or shape function). $\mathcal{B}(u,v)$ is
a bi-linear functional dependent upon the original and trial function  and $\mathcal{L}(v)$ a linear functional
dependent only of the trial function. The condition above is only possible to be obtained if the following
functional is minimized in the case of 1st order differential equations:

\begin{equation} \label{equation:functional}
\mathcal{V}(u,u) = \frac{1}{2} \mathcal{B}(u,u) - \mathcal{L}(u)
\end{equation} \\

In order to $u$ ($\frac{\partial \mathcal{V}}{\partial u} = 0$) using the shape function ($u = \mathbf{N}\mathbf{u}$),
holds the solution to $\mathcal{B}(u,v) = \mathcal{L}(v)$, and to the variational problem in eq. \ref{equation:variational}
\cite{Segerlind:1984,Rozanov:1998,Ghanem:1991}.

\subsubsection*{Shape Functions and Elements}

Shape function is the continuous approximate solution to the variational problem inside the finite element space.
The shape function is only dependent upon the type and shape of the finite element. Elements can be grouped into
their different interpolation functions: i) first order - linear elements; ii) second order - quadratic elements,
and iii) third order - cubic elements (higher order shape functions are unusual).

The most common FE shape functions are presented in Table \ref{table:FEShapeFunctions}. These describe a piecewise
solution to the variational problem in the physical domain. The FEM is generally used in the natural coordinate
system \cite{HenwoodandBonet:1996}. For example, the subtract concentration $C_S$ inside a linear rectangular element:

\begin{equation}
C_S(x,y) = N_iC_{S,i} + N_jC_{S,j} + N_kC_{S,k} + N_mC_{S,m}
\end{equation}

where $C_S(x,y)$ is the enzyme concentration in each of the physical positions inside the element, and
$C_{S,i}$, $C_{S,j}$, $C_{S,k}$, $C_{S,m}$ at the nodal presents $i$, $j$, $k$ and $m$ respectively,
$N_i$, $N_j$, $N_k$ and $N_m$ are the shape coefficients and are dependent on the elements coordinates
\cite{Segerlind:1984,Rozanov:1998,Ghanem:1991}.

The shape function can be presented in the compact matrix format:

\begin{equation}
    C_E(x,y)=
    \left[
    \begin{array}{lccl}
        N_i & N_j & N_k & N_m
    \end{array}
    \right] \cdot \left[
    \begin{array}{c}
        C_{E,i} \\
        C_{E,j} \\
        C_{E,k} \\
	C_{E,m}
    \end{array}
    \right] = \mathbf{N} \cdot \mathbf{C_E}
\end{equation}

where $N_i$ \ldots $N_m$ are the shape coefficients and $C_{E,i}$ \ldots $C_{E,m}$ the enzyme concentrations at
the element nodes $i$ to $m$, respectively. The discretization presented in this manuscript can be further extended to the
different types of elements using similar mathematical reasoning.

\subsection*{Reaction Models}

\subsubsection*{Reactions in space-time}

Diffusion dependent enzyme reactions are well described by the 2nd Fick law:

\begin{equation} \label{equation:2ndfick}
\frac{dC}{dt} - \nabla \left( D \nabla C \right) - r^* = f
\end{equation}

where $C$ is the specimen concentration ($mole.dm^{-3}$), $D$ the mass diffusivity ($m^2.s^{-1}$), $f$ the
force vector and $r^*$ the reaction rate of $C$ per unit value ($mole.s^{-1}.m^{-3}$). 
The simplest reaction term $r^*$ to be added to eq \ref{equation:2ndfick} is the first order kinetic:

\begin{center}
$A \rightharpoonup B$ 
\end{center} 

which can be described by:

\begin{equation}   \label{equation:1storder}
        \frac {dC_a}{dt} = - k C_a
\end{equation}

where the kinetic rate $k$ ($s^{-1}$) is a function of temperature given the modified Arrhenius law.
Such states that the decay of $A$ is proportional to the probability of finding $A$ ($p(A)$) molecule inside the
finite element space, that is $p(A) \propto C_a$. Consequently, the first order reaction is proportional to its
concentration inside the finite element. For the sake of simplicity, lets assume this reaction is occurring inside a
linear triangle where $C_(A)$ is given by:

\begin{equation}
C_a = N_iC_{a,i} + N_jC_{a,j} + N_kC_{a,k}
\end{equation} 

which is the space distribution of probabilities of finding ($A$) inside the finite element space.
Therefore, $N_i$, $N_j$ and $N_k$ map the random movements of $A$ molecules inside the finite elements,
proportional to speciemens concentration.

Once formulated the variational problem is possible to obtain:
\begin{eqnarray}
        V_\Omega = \int_\Omega \left( \frac {dC_a}{dt} + kC_a \right) \cdot v(\Omega) d\Omega \nonumber \\
        = \int_\Omega \frac {dC_a}{dt} \cdot d\Omega + \int_\Omega kC_a \cdot v(\Omega) d\Omega = 0
\end{eqnarray} \\

Under these circumstances the variational can be solved by using a linear
functional which holds the true solution after the minimization of the
bilinear functional $\mathcal{V}(C_a,v)$:

\begin{equation} 
        \mathcal{V}(C_a,v)=\frac {1}{2} \mathcal{B}(C_a,v) + \mathcal{L}(v)
\end{equation}

where the \textit{functionals} $\mathcal{B}(C_a,v)$ and $\mathcal{L}(v)$ are given by:

\begin{eqnarray} \nonumber
       \mathcal{B}(C_a,v) = \int_\Omega kC_a \cdot v(\Omega) d\Omega \\
       \mathcal{L}(v) = \int_\Omega \frac {dC_a}{dt} \cdot v(\Omega) d\Omega
\end{eqnarray} 

and therefore the functional $\mathcal{V}(C_a)$ takes the form of:

\begin{equation} \label{equation:1storderfunc}
       \mathcal{V}(C_a,v) = \int_\Omega \frac {1}{2} kC_a d\Omega
       + \int_\Omega \left( \frac {dC_a}{dt} \right) \cdot C_a d\Omega
\end{equation}

Substituting the \textit{element functions} in the first term of eq. \ref{equation:1storderfunc}, yields:

\begin{center}
\begin{eqnarray} \nonumber
        \mathcal{V} = \frac {1}{2} \int_\Omega \left( N_ik_i + N_jk_j + N_kk_k \right) \cdot \\
		\left( N_iC_{a,i} + N_jC_{a,j} + N_kC_{a,k} \right)^2 d\Omega 
\end{eqnarray}
\end{center}

That once minimised for the node $i$, holds:

\begin{center}
\begin{eqnarray}
        \frac {\partial \mathcal{V}}{\partial C_{a,i}} =
        \int_\Omega [
        \left( k_iN_i^3 + k_jN_i^2N_j + k_kN_i^2N_k \right) \cdot C_{a,i} \nonumber \\
        + \left( k_iN_i^2N_j + k_jN_iN_j^2 + k_kN_iN_jN_k \right) \cdot C_{a,j} \nonumber \\
        + \left( k_iN_i^2N_k + k_jN_iN_jN_k + k_kN_iN_k^2 \right) \cdot C_{a,k}
        ] \; \partial \Omega
\end{eqnarray} \\
\end{center}

The same minimization is necessary to be made in terms of $C_{a,j}$
and $C_{a,k}$, to obtain all the elements of the final matrix $\mathbf{K}$.
After algebraic manipulation, the stiffness matrix ($\mathbf{K}$) is possible
to be described in the matrix format by:

\begin{equation}
	\mathbf{K} = \int_\Omega \mathbf{N}^T \mathbf{k}^T \mathbf{N} \mathbf{N}^T  \partial \Omega
\end{equation} \\


where $k$ is the column vector $\mathbf{k}=\left[k_i~k_j~k_k\right]$, and the kinetic rate inside the
finite element is given by $\mathbf{Nk}$. If one considers $k$ as a row vector, than the solution is
$\mathbf{K} = \int_\Omega \mathbf{N}^T \mathbf{N} \mathbf{k}^t \mathbf{N}  \partial \Omega$, since these
are symmetric matrices. Similarly, for the second term of eq. \ref{equation:1storderfunc}:

\begin{center}
\begin{eqnarray} \nonumber
     \mathcal{V} =  \int_\Omega \left( N_i \frac{dC_{a,i}}{dt} + N_j \frac{dC_{a,j}}{dt}
     + N_k \frac{dC_{a,k}}{dt}\right) \cdot \\
   	\left( N_iC_{a,i} + N_jC_{a,j} + N_kC_{a,k} \right) \partial \Omega
\end{eqnarray}
\end{center}

That once minimised in terms of $C_{a,i}$ yields:

\begin{equation}
     \frac {\partial \mathcal{V}}{\partial C_{a,i}} =
     \int_\Omega \left( N_i^2 \frac{dC_{a,i}}{dt} + N_iN_j \frac{dC_{a,j}}{dt}
     + N_iN_k \frac{dC_{a,k}}{dt}\right)
      \partial \Omega
\end{equation}

The same kind of minimization is necessary for $C_{a,j}$ and $C_{a,k}$, to obtain
the final matrix that will enable the $FEM$ method computation.
After algebraic manipulation, the full minimization of the
variational is possible to be presented in the matrix format:

\begin{equation}
	\frac {\partial \mathcal{V}}{\partial C_a} = \int_\Omega \mathbf{N}^T \mathbf{N} \partial \Omega \cdot \mathbf{\dot{C}_a} 
\end{equation} 

Therefore, the chemical reaction can be computed accross the physical domain by:

\begin{equation} 
\int_\Omega \mathbf{N}^T \mathbf{N} \partial \Omega \cdot \mathbf{\dot{C}_a} + \int_\Omega \mathbf{K} \partial \Omega \cdot \mathbf{C_a} = 0
\end{equation}

where $\int_\Omega \mathbf{N}^T \mathbf{N} \partial\Omega$ presents the probabilities of random movements of the molecule $a$
in any direction inside the finite element and $\int_\Omega \mathbf{K} \partial\Omega$ the probabilities of effective
conformational changes of $a$ across the finite element.
An important assumption in this discretization, is the fact that kinetic rate is not constant across the
physical domain. Such occurs in non-homogeneous biological materials. If one considers constant kinetics,
than $\mathbf{K}= k_c \int_\Omega \mathbf{N}^T\mathbf{N}  \partial \Omega$, where $k_c$ is a constant kinetic rate.
For more chemical reaction mechanisms, please consult \cite{Martinsetal:2009b}.

Although the 1st order reaction kinetics is the most simplest mechanism, it is still the most widely used to represent
both systems chemistry and 'in-silico' organisms, where single steps are considered uni-molecular, and in the last case,
catalyzed by an enzyme, being possible to be used in conjunction with reaction networks and GSM.

\subsubsection*{Second order kinetics} \label{section:2ndorder}

The simplest form of reaction given by molecular colisions, is the second order reaction kinetic:

\begin{center}
	$A + B \rightharpoonup C$
\end{center}

where,

\begin{equation}
\frac{dC_a}{dt} = -kC_aC_b
\end{equation} \\

$C_a$ and $C_b$ are the concentrations of $A$ and $B$ specimens inside the finite element. 
In this case, reaction only occurs once there are effective collisions between $A$ and $B$.
Therefore, inside any linear finite element the variational form is presented as follows:

\begin{equation}
 \mathcal{V} = \int_\Omega \frac{dCa}{dt}  v(\Omega) d\Omega + \frac{1}{2} \int_\Omega k C_aC_b v(\Omega) d\Omega
\end{equation}

where $k$, $C_a$ and $C_b$ vary consistently inside the finite element, taking the form:

\begin{equation}
 \mathcal{V}(C_a) = \int_\Omega \frac{dCa}{dt} C_a d\Omega + \frac{1}{2} \int_\Omega k C_bC_a^2 d\Omega
\end{equation}

which minimizing for node $i$, $j$ and $k$, attains:

\begin{equation}
 \frac{\partial \mathcal{V}}{\partial C_{a,i}} = \int_\Omega \frac{dCa}{dt}N_i d\Omega +  \int_\Omega k C_bC_aN_i d\Omega
\end{equation}

Which after the variational minimization, the solution yields:

\begin{equation}
\int_\Omega \mathbf{N}^T \mathbf{N} d\Omega \cdot \mathbf{\dot{C}}_a + \int_\Omega \mathbf{M} d\Omega \cdot \mathbf{C}_a\mathbf{C}_b^T \cdot \mathbf{U}= 0
\end{equation}

where, $\mathbf{M} = diag(\mathbf{N}) \mathbf{K}$, and $\mathbf{U}$ is the column vector $\left[ 1~1~1 \right]$.
$\mathbf{M}$ expresses the frequency of $A$ and $B$ to react inside the finite element. Afterwards, both 
equations for $C_a$ and $C_b$ solution must be computed with both equations.
Moreover, the term $C_aC_b^T$ expresses all possible collision probabilities between $a$ and $b$ specimens
inside the finite element. The same is possible to derive for auto-catalyzed reactions ($A + A \rightharpoonup C$),
being possible to show that the solution is held by:
$\int_\Omega \mathbf{N}^T \mathbf{N} d\Omega \cdot \mathbf{\dot{C}}_a + \int_\Omega \mathbf{M} d\Omega \cdot \mathbf{C}_a\mathbf{C}_a^T \cdot \mathbf{U}= 0$
.

\section*{Results and Discussion}

\subsection*{Enzymatic models}

\subsubsection*{Enzyme activation}

The enzymatic activation/inactivation is an example of fractional conversion model
\cite{VillotaandHawkes:1992}, that describes an equilibrium between two species $E^0$ and $E$,
which correspond to inactive and active enzymes, respectively.

\begin{center}
	$E^0 \rightleftharpoons E$ \\
\end{center}

In this case, the concentration of $E^0$ and $E$ is established by a dynamical equilibrium by:

\begin{eqnarray}
\frac{dC_{E^0}}{dt} + k_1C_{E^0} - k_{-1}C_E = 0 \\
\frac{dC_E}{dt} - k_1C_{E^0} + k_{-1}C_E = 0
\end{eqnarray} \\

After manipulations, the following finite element formulation inside the linear finite element for inactive
and active enzymes, respectively:

\begin{equation} \nonumber
	\int_\Omega \mathbf{N^T} \mathbf{N} \partial\Omega \cdot \mathbf{\dot C_{E^0}}
		+ \int_\Omega \mathbf{K_1} \partial\Omega \cdot {C_{E^0}} \\
			- \int_\Omega \mathbf{K_{-1}} \partial\Omega \cdot {C_E} = 0
\end{equation}

\subsubsection*{Reaction chain}

Consider the following reaction chain:

\begin{center}
	$\ldots \rightharpoonup A_{i-1} \rightharpoonup A_{i} \rightharpoonup A_{i+1} \rightharpoonup \ldots$
\end{center}

where $i$ is the i'th specimen in the reaction chain. By direct comparison with previous formulations it is
simple to derive the FEM formulation for each specimen:

\begin{equation} \nonumber
\int_\Omega \mathbf{N^T} \mathbf{N} \partial\Omega \cdot \mathbf{\dot{C}a_i}
- \int_\Omega \mathbf{K_i} \partial\Omega \cdot {Ca_{i} } \\
+ \int_\Omega \mathbf{K_{i-1}} \partial\Omega \cdot Ca_{i-1} = 0
\end{equation}

The full reaction chain is computed by joining all the specimens equation matrixes.

\subsubsection*{Michaelis-Menten model}

The most widely known enzymatic model is the Michaelis-Menten mechanism: \\

\begin{center}
$E + S \rightleftharpoons ES\rightharpoonup E + P$	\\
\end{center}

which can be expressed by a balance to each species:

\begin{equation}
\frac{dC_{ES}}{dt} - k_1C_EC_S + k_{-1}C_{ES} + k_2C_{ES} = 0\\
\end{equation}

\begin{equation}
\frac{dC_S}{dt} + k_1C_EC_S - k_{-1}C_{ES} = 0\\
\end{equation}

\begin{equation}
\frac{dC_P}{dt} - k_2C_{ES} = 0\\
\end{equation}

\begin{equation}
\frac{dC_E}{dt} + k_1C_EC_S - k_{-1}C_{ES} - k_2C_{ES} = 0
\end{equation}

which inside the finite element can be expressed as:

\begin{center}
\begin{eqnarray} \nonumber
\int_\Omega\mathbf{N}^T \mathbf{N} \partial \Omega \cdot  \mathbf{\dot C_{ES}}
- \int_\Omega \mathbf{M}_1 \partial\Omega \cdot \mathbf{C_{ES}} \cdot \mathbf{C_E^T} \cdot \mathbf{U} \\ 
+ \int_\Omega \mathbf{K_{-1}} \partial \Omega \cdot \mathbf{C_{ES}}
+ \int_\Omega \mathbf{K_2} \partial \Omega \cdot \mathbf{C_{ES}} = 0
\end{eqnarray}		
\end{center}

\begin{center}
\begin{eqnarray} \nonumber
\int_\Omega \mathbf{N}^T \mathbf{N} \partial \Omega \cdot \dot{\mathbf{C_S}}
 + \int_\Omega \mathbf{M} \partial\Omega \cdot \mathbf{C_E} \cdot \mathbf{C_S^T} \cdot \mathbf{U} \\
 - \int_\Omega \mathbf{K_{-1}} \partial \Omega \cdot \mathbf{C_{ES}} = 0
\end{eqnarray}
\end{center}

\begin{center}
\begin{equation} \nonumber
\int_\Omega \mathbf{N}^T \mathbf{N} \partial \Omega \cdot \mathbf{C_P}
- \int_\Omega \mathbf{K_2} \partial\Omega \cdot \mathbf{C_{ES}} = 0
\end{equation}
\end{center}

\begin{center}
\begin{eqnarray} \nonumber
\int_\Omega\mathbf{N}^T \mathbf{N}\partial\Omega \cdot \dot{\mathbf{C_E}}
+ \int_\Omega \mathbf{M} \partial\Omega \cdot \mathbf{C_E} \cdot \mathbf{C_S^T} \cdot \mathbf{U} \\
- \int_\Omega \mathbf{K_{-1}} \cdot \mathbf{C_{ES}} - \int_\Omega \mathbf{K_2} \partial\Omega \cdot \mathbf{C_{ES}} = 0
\end{eqnarray}
\end{center}

\subsubsection*{Inhibition of enzymatic activity}

Enzyme inactivation is both a control mechanisms, as well as, a lowering yield factor
by an inhibitor $I$ (natural or synthetic) which interacts with the enzyme,
decreasing the catalytic activity. 

\subsubsection*{Competitive inhibition}

\begin{figure} 
        \begin{center}
           \vspace{1cm}
	   \includegraphics[scale=0.8] {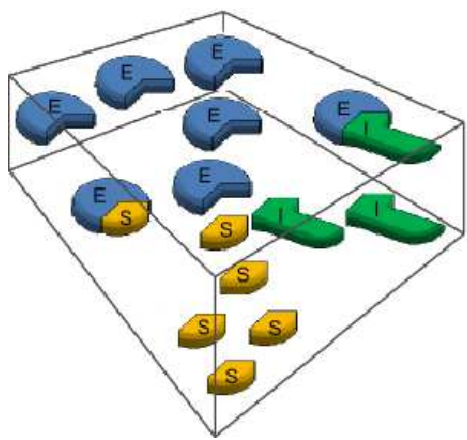}
	   \caption{Competitive inhibition inside a 3D finite element: reaction rates are function of local
concentration of specimens and temperature.}
           \label{figure:ExperimentalDiagram}
        \end{center}
\end{figure}

Competitive inhibition occurs when an inhibitor ($I$) (Figure \ref{figure:ExperimentalDiagram})
competes with the substrate for the active center, being represented by:

\begin{center}
	$E + S  \rightleftharpoons ES \rightharpoonup E + P$ \\
\end{center}
\begin{center}
	$E + I  \rightleftharpoons EI$ \\ 
\end{center}

which can be expressed by a balance to each species:

\begin{equation}
\frac{dC_{ES}}{dt} - k_1C_EC_S + k_{-1}C_{ES} + k_2C_{ES} = 0
\end{equation}

\begin{equation}
\frac{dC_S}{dt} + k_1C_EC_S - k_{-1}C_{ES} = 0
\end{equation}

\begin{equation}
\frac{dC_P}{dt} - k_2C_{ES} = 0
\end{equation}

\begin{equation}
\frac{dC_E}{dt} + k_1C_EC_S - k_{-1}C_{ES} - k_2C_{ES} + k_3C_EC_I - k_{-3}C_{EI} = 0
\end{equation}

\begin{equation}
\frac{dC_{EI}}{dt} - k_3C_EC_I + k_{-3}C_{EI} = 0
\end{equation}

\begin{equation}
\frac{dI}{dt} + k_IC_EC_I - k_{-I}C_{EI} = 0
\end{equation} \\

and after solving the variational problem, the solution can be expressed as:

\begin{center}
\begin{eqnarray} \nonumber 
\int_\Omega\mathbf{N}^T \mathbf{N} \partial \Omega \cdot \mathbf{\dot C_{ES}} \\ \nonumber
- \int_\Omega \mathbf{M}_1 \partial\Omega \cdot \mathbf{C_E} \cdot \mathbf{C_S^T} \cdot \mathbf{U} \\
+ \int_\Omega \mathbf{K_{-1}} \partial \Omega \cdot \mathbf{C_{ES}}
+ \int_\Omega \mathbf{K_2} \partial \Omega \cdot \mathbf{C_{ES}} = 0
\end{eqnarray}		
\end{center}

\begin{center}
\begin{eqnarray} \nonumber
\int_\Omega\mathbf{N}^T \mathbf{N}\partial\Omega \cdot \mathbf{C_S}
+ \int_\Omega \mathbf{M}_1 \partial\Omega \cdot \mathbf{C_E} \cdot \mathbf{C_S^T} \cdot \mathbf{U} \\
- \int_\Omega \mathbf{K_{-1}} \partial\Omega \cdot \mathbf{C_{ES}} = 0
\end{eqnarray}
\end{center}

\begin{center}
\begin{equation}
\int_\Omega \mathbf{N}^T \mathbf{N} \partial\Omega \cdot \mathbf{C_P} - \int_\Omega \mathbf{K_2} \partial\Omega \cdot \mathbf{C_{ES}} = 0
\end{equation}
\end{center}

\begin{center}
\begin{eqnarray} \nonumber
\int_\Omega\mathbf{N}^T \mathbf{N}\partial\Omega \cdot  \mathbf{\dot C_E}
+ \int_\Omega \mathbf{M}_1 \partial\Omega \cdot \mathbf{C_E} \cdot \mathbf{C_S^T} \cdot \mathbf{U} \\ \nonumber
- \int_\Omega \mathbf{K_{-1}} \partial\Omega \cdot \mathbf{C_{ES}}
- \int_\Omega \mathbf{K_2} \partial\Omega \cdot \mathbf{C_{ES}} \\
+ \int_\Omega \mathbf{M}_2 \partial\Omega \cdot \mathbf{C_E} \cdot \mathbf{C_I^T} \cdot \mathbf{U} 
- \int_\Omega \mathbf{K_{-3}} \partial\Omega \cdot \mathbf{C_{EI}} = 0
\end{eqnarray} 
\end{center}

\begin{center}
\begin{eqnarray} \nonumber
\int_\Omega\mathbf{N}^T \mathbf{N}\partial\Omega \cdot \mathbf{C_{EI}}
- \int_\Omega \mathbf{M}_2 \partial\Omega \cdot \mathbf{C_E} \cdot \mathbf{C_I^T} \cdot \mathbf{U} \\
+ \int_\Omega \mathbf{K_{-3}} \partial\Omega \cdot \mathbf{C_{EI}} = 0
\end{eqnarray}
\end{center}

\begin{center}
\begin{eqnarray} \nonumber
\int_\Omega\mathbf{N}^T \mathbf{N}\partial\Omega \cdot \mathbf{C_I}
+ \int_\Omega \mathbf{M}_2 \partial\Omega \cdot \mathbf{C_E} \cdot \mathbf{C_I^T} \cdot \mathbf{U} \\
- \int_\Omega \mathbf{K_{-3}} \partial\Omega \cdot \mathbf{C_{EI}} = 0 
\end{eqnarray}
\end{center}

where $\mathbf{M}_1$ and $\mathbf{M}_2$ express the frequency of E-S and E-I to react inside the finite element.

\subsubsection*{Non-competitive inhibition}

Non-competitive inhibition occurs when an inhibitor ($I$) reversibly establishes a chemical bound
with the enzyme which is not the active site, but nevertheless affects its catalytic activity,
being possible to be expressed by the mechanism:
	
\begin{center}
	$E + S  \rightleftharpoons  ES \rightharpoonup E + P$ \\
\end{center}
	
\begin{center}
	$E + I  \rightleftharpoons  EI$ \\
\end{center}

\begin{center}
	$ES + I \rightleftharpoons  ESI$ \\
\end{center}

\begin{center}
	$EI + S \rightleftharpoons  ESI$ \\
\end{center}

which can be expressed by a balance to each species:

\begin{equation}
\frac{dC_{ES}}{dt} - k_1C_EC_S + k_{-1}C_{ES} + k_2C_{ES} + k_4C_{ES}C_I - k_{-4}C_{ESI}= 0
\end{equation}

\begin{equation}
\frac{dC_S}{dt} + k_1C_EC_S - k_{-1}C_{ES} - k_{-5}C_{ESI} + k_5C_{EI}C_{S} = 0
\end{equation}

\begin{equation}
\frac{dC_P}{dt} - k_2C_{ES} = 0
\end{equation}

\begin{equation}
\frac{dC_E}{dt} + k_1C_ECO_S - k_{-1}C_{ES} - k_2C_{ES} + k_3C_EC_I - k_{-3}C_{EI} = 0
\end{equation}

\begin{equation}
\frac{dC_{EI}}{dt} - k_3C_EC_I + k_{-3}C_{EI}C_S + k_5C_{EI}C_S - k_{-5}C_{ESI} = 0
\end{equation}

\begin{equation}
\frac{dC_{ESI}}{dt} + k_{-5}C_{ESI} - k_5C_{EI}C_{S} + k_{-4}C_{ESI} - k_{4}C_{ES}C_{I} = 0
\end{equation}

\begin{equation}
\frac{dC_I}{dt} - k_{-4}C_{ESI} + k_4C_{ES}C_I - k_{-3}C_{EI} + k_3C_EC_I = 0
\end{equation}

which inside the finite element can be expressed as:

\begin{center}
\begin{eqnarray} \nonumber
\int_\Omega\mathbf{N}^T \mathbf{N} \partial \Omega \cdot \mathbf{C_{ES}}
- \int_\Omega \mathbf{M}_1 \partial\Omega \cdot \mathbf{C_E} \cdot \mathbf{C_S^T} \cdot \mathbf{U} \\ \nonumber
+ \int_\Omega \mathbf{K_{-1}} \partial \Omega \cdot \mathbf{C_{ES}}
+ \int_\Omega \mathbf{K_2} \partial \Omega \cdot \mathbf{C_{ES}} \\
+ \int_\Omega \mathbf{M}_2 \partial\Omega \cdot \mathbf{C_{ES}} \cdot \mathbf{C_I^T} \cdot \mathbf{U}
- \int_\Omega \mathbf{K_{-4}} \partial \Omega \cdot \mathbf{C_{ESI}} = 0 
\end{eqnarray}		
\end{center}

\begin{center}
\begin{eqnarray} \nonumber
\int_\Omega\mathbf{N}^T \mathbf{N}\partial\Omega \cdot \mathbf{C_S} 
+ \int_\Omega \mathbf{M}_1 \partial\Omega \cdot \mathbf{C_E} \cdot \mathbf{C_S^T} \cdot \mathbf{U} \\ \nonumber
- \int_\Omega \mathbf{K_{-1}} \partial\Omega \cdot \mathbf{C_{ES}}
- \int_\Omega \mathbf{K_{-5}} \partial\Omega \cdot \mathbf{C_{ESI}} \\
+ \int_\Omega \mathbf{M}_2 \partial\Omega \cdot \mathbf{C_{EI}} \cdot \mathbf{C_S^T} \cdot \mathbf{U} = 0
\end{eqnarray}
\end{center}

\begin{center}
\begin{equation}
\int_\Omega\mathbf{N}^T \mathbf{N}\partial\Omega \cdot \mathbf{C_P} - \int_\Omega \mathbf{K_2} \partial\Omega \cdot \mathbf{C_{ES}} = 0
\end{equation}
\end{center}

\begin{center}
\begin{eqnarray} \nonumber
\int_\Omega\mathbf{N}^T \mathbf{N}\partial\Omega \cdot \mathbf{C_E}
+ \int_\Omega \mathbf{M}_1 \partial\Omega \cdot \mathbf{C_E} \cdot \mathbf{C_S^T} \cdot \mathbf{U} \\ \nonumber
- \int_\Omega \mathbf{K_{-1}} \cdot \mathbf{C_{ES}} - \int_\Omega \mathbf{K_2} \partial\Omega \cdot \mathbf{C_{ES}} \\
+ \int_\Omega \mathbf{M}_3 \partial\Omega \cdot \mathbf{C_E} \cdot \mathbf{C_I^T} \cdot \mathbf{U} - \int_\Omega \mathbf{K_{-3}} \partial\Omega \cdot \mathbf{C_{EI}} = 0
\end{eqnarray} 
\end{center}

\begin{center}
\begin{eqnarray} \nonumber
\int_\Omega\mathbf{N}^T \mathbf{N}\partial\Omega \cdot \mathbf{C_{EI}}
- \int_\Omega \mathbf{M}_1 \partial\Omega \cdot \mathbf{C_E} \cdot \mathbf{C_I^T} \cdot \mathbf{U} \\ \nonumber
+ \int_\Omega \mathbf{M}_2 \partial\Omega \cdot \mathbf{C_{EI}} \cdot \mathbf{C_S^T} \cdot \mathbf{U}
+ \int_\Omega \mathbf{M}_4 \partial\Omega \cdot \mathbf{C_{EI}} \cdot \mathbf{C_S^T} \cdot \mathbf{U} \\
- \int_\Omega\mathbf{K_{-5}} \partial\Omega \cdot \mathbf{C_{ESI}} = 0
\end{eqnarray}
\end{center}

\begin{center}
\begin{eqnarray} \nonumber
\int_\Omega\mathbf{N}^T \mathbf{N}\partial\Omega \cdot \mathbf{C_{ESI}}
+ \int_\Omega\mathbf{K_{-5}} \partial\Omega \cdot \mathbf{C_{ESI}}\\ \nonumber
- \int_\Omega \mathbf{M}_4 \partial\Omega \cdot \mathbf{C_{EI}} \cdot \mathbf{C_S^T} \cdot \mathbf{U}
+ \int_\Omega\mathbf{K_{-4}} \partial\Omega \cdot \mathbf{C_{ESI}} \\
- \int_\Omega \mathbf{M}_4 \partial\Omega \cdot \mathbf{C_{ES}} \cdot \mathbf{C_I^T} \cdot \mathbf{U} = 0
\end{eqnarray}
\end{center}

\begin{center}
\begin{eqnarray} \nonumber
\int_\Omega\mathbf{N}^T \mathbf{N}\partial\Omega \cdot \mathbf{C_I}
- \int_\Omega\mathbf{K_{-4}} \partial\Omega \cdot \mathbf{C_{ESI}} \\ \nonumber
+ \int_\Omega \mathbf{M}_4 \partial\Omega \cdot \mathbf{C_{ES}} \cdot \mathbf{C_I^T} \cdot \mathbf{U}
- \int_\Omega \mathbf{K_{-3}} \partial\Omega \cdot \mathbf{C_{EI}} \\
+ \int_\Omega \mathbf{M}_3 \partial\Omega \cdot \mathbf{C_E} \cdot \mathbf{C_I^T} \cdot \mathbf{U} = 0
\end{eqnarray}
\end{center}

where $\mathbf{M}_1$, $\mathbf{M}_2$, $\mathbf{M}_3$, $\mathbf{M}_4$ express colliding probabilities of
$E-S$, $ES-I$, $E-I$ and $EI-S$.

\subsubsection*{Anti-competitive inhibition}

When an inhibitor links itself reversibly to enzyme-substrate complex and not to the free enzyme,
this is known as anti-competitive inhibition mechanism:

\begin{center}
	$E + S  \rightleftharpoons  ES \rightharpoonup E + P$ \\
\end{center}
\begin{center}
	$ES + I  \rightleftharpoons  ESI$
\end{center}

which can be expressed by a balance to each species:

\begin{equation}
\frac{dC_{ES}}{dt} - k_1C_EC_S + k_{-1}C_{ES} + k_2C_{ES} \\
+ k_3C_{ES}C_I - k_{-3}C_{ESI} = 0 
\end{equation}

\begin{equation}
\frac{dC_S}{dt} + k_1C_EC_S - k_{-1}C_{ES} = 0
\end{equation}

\begin{equation}
\frac{dC_P}{dt} - k_2C_{ES} = 0
\end{equation}

\begin{equation}
\frac{dC_E}{dt} + k_1C_EC_S - k_{-1}C_{ES} - k_2C_{ES} = 0
\end{equation}

\begin{equation}
\frac{dC_{ESI}}{dt} - k_3C_{ES}C_I + k_{-3}C_{ESI} = 0
\end{equation}

\begin{equation}
\frac{dC_I}{dt} + k_{3}C_{ES}C_I - k_{-3}C_{ESI} = 0
\end{equation}

which inside the finite element can be expressed as:

\begin{center}
\begin{eqnarray} \nonumber
\int_\Omega\mathbf{N}^T \mathbf{N} \partial \Omega \cdot \mathbf{C_{ES}}
- \int_\Omega \mathbf{M}_1 \partial\Omega \cdot \mathbf{C_E} \cdot \mathbf{C_S^T} \cdot \mathbf{U} \\ \nonumber
+ \int_\Omega \mathbf{K_{-1}} \partial \Omega \cdot \mathbf{C_{ES}}
+ \int_\Omega \mathbf{K_2} \partial \Omega \cdot \mathbf{C_{ES}} \\
+ \int_\Omega \mathbf{M}_2 \partial\Omega \cdot \mathbf{C_{ES}} \cdot \mathbf{C_I^T} \cdot \mathbf{U}
- \int_\Omega \mathbf{K_{-3}} \partial \Omega \cdot \mathbf{C_{ESI}} = 0
\end{eqnarray}		
\end{center}

\begin{center}
\begin{eqnarray} \nonumber
\int_\Omega\mathbf{N}^T \mathbf{N}\partial\Omega \cdot \mathbf{C_S}
+ \int_\Omega \mathbf{M}_1 \partial\Omega \cdot \mathbf{C_E^T} \cdot \mathbf{C_S^T} \cdot \mathbf{U} \\
- \int_\Omega \mathbf{K_{-1}} \partial\Omega \cdot \mathbf{C_{ES}} = 0
\end{eqnarray}
\end{center}

\begin{center}
\begin{eqnarray}
\int_\Omega\mathbf{N}^T \mathbf{N}\partial\Omega \cdot  \mathbf{\dot C_P}
- \int_\Omega \mathbf{K_2} \partial\Omega \cdot \mathbf{C_{ES}} = 0
\end{eqnarray}
\end{center}

\begin{center}
\begin{eqnarray} \nonumber
\int_\Omega\mathbf{N}^T \mathbf{N}\partial\Omega \cdot  \mathbf{\dot C_E}
+ \int_\Omega \mathbf{M}_2 \partial\Omega \cdot \mathbf{C_E} \cdot \mathbf{C_S^T} \cdot \mathbf{U} \\
- \int_\Omega \mathbf{K_{-1}} \partial\Omega \cdot \mathbf{C_{ES}}
- \int_\Omega \mathbf{K_2} \partial\Omega \cdot \mathbf{C_{ES}} = 0
\end{eqnarray} 
\end{center}

\begin{center}
\begin{eqnarray} \nonumber
 \int_\Omega\mathbf{N}^T \mathbf{N}\partial\Omega \cdot \mathbf{C_{ESI}}
- \int_\Omega \mathbf{M}_1 \partial\Omega \cdot \mathbf{C_{ES}} \cdot \mathbf{C_I^T} \cdot \mathbf{U} \\
+ \int_\Omega\mathbf{K_{-3}} \partial\Omega \cdot \mathbf{C_{ESI}} = 0
\end{eqnarray}
\end{center}

\begin{center}
\begin{eqnarray} \nonumber
\int_\Omega\mathbf{N}^T \mathbf{N}\partial\Omega \cdot  \mathbf{\dot C_I}
+ \int_\Omega \mathbf{M}_2 \partial\Omega \cdot \mathbf{C_{ES}} \cdot \mathbf{C_I^T} \cdot \mathbf{U} \\
- \int_\Omega \mathbf{K_{-3}} \partial\Omega \cdot \mathbf{C_{ESI}} = 0
\end{eqnarray}
\end{center}

where $\mathbf{M}_1$ and $\mathbf{M}_2$ express the frequency of $E-S$ and $ES-I$ colisions
the finite element.

\subsubsection*{Ping-Pong Bi-Bi mechanism}

In Ping-Pong Bi-Bi mechanisms, one of the substrates connects to the enzyme and one of the resulting
products releases before the second substrate can connect:

\begin{center}
	$E + A  \rightharpoonup  E A \rightleftharpoons E^* P \rightharpoonup E^* + P$
\end{center}
\begin{center}
	$E^* + B \rightharpoonup E^* B \rightleftharpoons E Q \rightharpoonup E + Q$
\end{center}						
		
which can be expressed by a balance to each species:

\begin{center}
\begin{eqnarray} \nonumber
\int_\Omega\mathbf{N}^T \mathbf{N} \partial \Omega \cdot   \mathbf{\dot C_E}
+ \int_\Omega \mathbf{M}_1 \partial\Omega \cdot \mathbf{C_E} \cdot \mathbf{C_A^T} \cdot \mathbf{U}\\
- \int_\Omega \mathbf{K_6} \partial \Omega \cdot \mathbf{C_{EQ}} = 0 
\end{eqnarray}
\end{center}

\begin{center}
\begin{eqnarray}
\int_\Omega\mathbf{N}^T \mathbf{N}\partial\Omega \cdot  \mathbf{\dot C_A}
	+ \int_\Omega \mathbf{M}_2 \partial\Omega \cdot \mathbf{C_E} \cdot \mathbf{C_S^T} \cdot \mathbf{U} = 0
\end{eqnarray}
\end{center}

\begin{center}
\begin{eqnarray} \nonumber
\int_\Omega\mathbf{N}^T \mathbf{N}\partial\Omega \cdot  \mathbf{\dot C_{EA}}
	- \int_\Omega \mathbf{M}_1 \partial\Omega \cdot \mathbf{C_E} \cdot \mathbf{C_A^T} \cdot \mathbf{U} \\
		- \int_\Omega \mathbf{K_{-2}} \partial\Omega \cdot \mathbf{C_{E^*P}}
		+ \int_\Omega \mathbf{K_2} \partial\Omega \cdot \mathbf{C_{EA}} = 0
\end{eqnarray}
\end{center}

\begin{center}
\begin{eqnarray} \nonumber
\int_\Omega\mathbf{N}^T \mathbf{N}\partial\Omega \cdot  \mathbf{\dot C_{E^*}}
- \int_\Omega \mathbf{K_3} \partial\Omega \cdot \mathbf{C_{E^*P}} \\
+ \int_\Omega \mathbf{M}_3 \partial\Omega \cdot \mathbf{C_{E^*}} \cdot \mathbf{C_B^T} \cdot \mathbf{U} = 0
\end{eqnarray} 
\end{center}

\begin{center}
\begin{eqnarray} \nonumber
\int_\Omega\mathbf{N}^T \mathbf{N}\partial\Omega \cdot  \mathbf{\dot C_{E^*P}}
- \int_\Omega \mathbf{K_{-2}} \partial\Omega \cdot \mathbf{C_{E^*P}} \\
+ \int_\Omega \mathbf{K_3} \partial\Omega \cdot \mathbf{C_{E^*P}} = 0
\end{eqnarray}
\end{center}

\begin{center}
\begin{eqnarray}
\int_\Omega\mathbf{N}^T \mathbf{N}\partial\Omega \cdot  \mathbf{\dot C_P}
- \int_\Omega \mathbf{K_3} \partial\Omega \cdot \mathbf{C_{E^*P}} = 0
\end{eqnarray}
\end{center}
\begin{center}
\begin{eqnarray} \nonumber
\int_\Omega\mathbf{N}^T \mathbf{N} \partial \Omega \cdot   \mathbf{\dot C_E}
+ \int_\Omega \mathbf{M}_1 \partial\Omega \cdot \mathbf{C_E} \cdot \mathbf{C_A^T} \cdot \mathbf{U}\\
- \int_\Omega \mathbf{K_6} \partial \Omega \cdot \mathbf{C_{EQ}} = 0 
\end{eqnarray}
\end{center}

\begin{center}
\begin{eqnarray}
\int_\Omega\mathbf{N}^T \mathbf{N}\partial\Omega \cdot  \mathbf{\dot C_A}
	+ \int_\Omega \mathbf{M}_2 \partial\Omega \cdot \mathbf{C_E} \cdot \mathbf{C_S^T} \cdot \mathbf{U} = 0
\end{eqnarray}
\end{center}

\begin{center}
\begin{eqnarray} \nonumber
\int_\Omega\mathbf{N}^T \mathbf{N}\partial\Omega \cdot  \mathbf{\dot C_{EA}}
	- \int_\Omega \mathbf{M}_1 \partial\Omega \cdot \mathbf{C_E} \cdot \mathbf{C_A^T} \cdot \mathbf{U} \\
		- \int_\Omega \mathbf{K_{-2}} \partial\Omega \cdot \mathbf{C_{E^*P}}
		+ \int_\Omega \mathbf{K_2} \partial\Omega \cdot \mathbf{C_{EA}} = 0
\end{eqnarray}
\end{center}

\begin{center}
\begin{eqnarray} \nonumber
\int_\Omega\mathbf{N}^T \mathbf{N}\partial\Omega \cdot  \mathbf{\dot C_{E^*}}
- \int_\Omega \mathbf{K_3} \partial\Omega \cdot \mathbf{C_{E^*P}} \\
+ \int_\Omega \mathbf{M}_3 \partial\Omega \cdot \mathbf{C_{E^*}} \cdot \mathbf{C_B^T} \cdot \mathbf{U} = 0
\end{eqnarray} 
\end{center}

\begin{center}
\begin{eqnarray} \nonumber
\int_\Omega\mathbf{N}^T \mathbf{N}\partial\Omega \cdot  \mathbf{\dot C_{E^*P}}
- \int_\Omega \mathbf{K_{-2}} \partial\Omega \cdot \mathbf{C_{E^*P}} \\
+ \int_\Omega \mathbf{K_3} \partial\Omega \cdot \mathbf{C_{E^*P}} = 0
\end{eqnarray}
\end{center}

\begin{center}
\begin{eqnarray}
\int_\Omega\mathbf{N}^T \mathbf{N}\partial\Omega \cdot  \mathbf{\dot C_P}
- \int_\Omega \mathbf{K_3} \partial\Omega \cdot \mathbf{C_{E^*P}} = 0
\end{eqnarray}
\end{center}

\begin{center}
\begin{eqnarray}
\int_\Omega\mathbf{N}^T \mathbf{N}\partial\Omega \cdot  \mathbf{\dot C_B}
+ \int_\Omega \mathbf{M}_3 \partial\Omega \cdot \mathbf{C_{E^*}} \cdot \mathbf{C_B^T} \cdot \mathbf{U} = 0
\end{eqnarray}
\end{center}

\begin{center}
\begin{eqnarray} \nonumber
\int_\Omega\mathbf{N}^T \mathbf{N}\partial\Omega \cdot  \mathbf{\dot C_{E^*B}}
- \int_\Omega \mathbf{M}_3 \partial\Omega \cdot \mathbf{C_{E^*}} \cdot \mathbf{C_B^T} \cdot \mathbf{U} \\ \nonumber
+ \int_\Omega \mathbf{K_5} \partial\Omega \cdot \mathbf{C_{E^*B}}
+ \int_\Omega \mathbf{K_{-5}} \partial\Omega \cdot \mathbf{C_{EQ}} \\
\int_\Omega \mathbf{K_6} \partial\Omega \cdot \mathbf{C_{EQ}}  = 0
\end{eqnarray}
\end{center}

\begin{center}
\begin{eqnarray} \nonumber
\int_\Omega\mathbf{N}^T \mathbf{N}\partial\Omega \cdot \mathbf{C_{EQ}}
- \int_\Omega \mathbf{K_5} \partial\Omega \cdot \mathbf{C_{E*B}} \\
+ \int_\Omega \mathbf{K_{-5}} \partial\Omega \cdot \mathbf{C_{EQ}}
+  \int_\Omega \mathbf{K_6} \partial\Omega \cdot \mathbf{C_{EQ}} = 0
\end{eqnarray}
\end{center}

\begin{center}
\begin{eqnarray}
\int_\Omega\mathbf{N}^T \mathbf{N}\partial\Omega \cdot  \mathbf{\dot C_Q}
- \int_\Omega \mathbf{K_6} \partial\Omega \cdot \mathbf{C_{EQ}} = 0
\end{eqnarray}
\end{center}

where $\mathbf{M}_1$, $\mathbf{M}_2$, $\mathbf{M}_3$ expresses the collision probabilities
of $E-A$, $E-S$, and $E-B$, respectively.

\begin{center}
\begin{eqnarray}
\int_\Omega\mathbf{N}^T \mathbf{N}\partial\Omega \cdot  \mathbf{\dot C_B}
+ \int_\Omega \mathbf{M}_3 \partial\Omega \cdot \mathbf{C_{E^*}} \cdot \mathbf{C_B^T} \cdot \mathbf{U} = 0
\end{eqnarray}
\end{center}

\begin{center}
\begin{eqnarray} \nonumber
\int_\Omega\mathbf{N}^T \mathbf{N}\partial\Omega \cdot  \mathbf{\dot C_{E^*B}}
- \int_\Omega \mathbf{M}_3 \partial\Omega \cdot \mathbf{C_{E^*}} \cdot \mathbf{C_B^T} \cdot \mathbf{U} \\ \nonumber
+ \int_\Omega \mathbf{K_5} \partial\Omega \cdot \mathbf{C_{E^*B}}
+ \int_\Omega \mathbf{K_{-5}} \partial\Omega \cdot \mathbf{C_{EQ}} \\
\int_\Omega \mathbf{K_6} \partial\Omega \cdot \mathbf{C_{EQ}}  = 0
\end{eqnarray}
\end{center}

\begin{center}
\begin{eqnarray} \nonumber
\int_\Omega\mathbf{N}^T \mathbf{N}\partial\Omega \cdot \mathbf{C_{EQ}}
- \int_\Omega \mathbf{K_5} \partial\Omega \cdot \mathbf{C_{E*B}} \\
+ \int_\Omega \mathbf{K_{-5}} \partial\Omega \cdot \mathbf{C_{EQ}}
+  \int_\Omega \mathbf{K_6} \partial\Omega \cdot \mathbf{C_{EQ}} = 0
\end{eqnarray}
\end{center}

\begin{center}
\begin{eqnarray}
\int_\Omega\mathbf{N}^T \mathbf{N}\partial\Omega \cdot  \mathbf{\dot C_Q}
- \int_\Omega \mathbf{K_6} \partial\Omega \cdot \mathbf{C_{EQ}} = 0
\end{eqnarray}
\end{center}

where $\mathbf{M}_1$, $\mathbf{M}_2$, $\mathbf{M}_3$ expresses the collision probabilities
of $E-A$, $E-S$, and $E-B$, respectively.
\begin{center}
\begin{eqnarray} \nonumber
\int_\Omega\mathbf{N}^T \mathbf{N} \partial \Omega \cdot   \mathbf{\dot C_E}
+ \int_\Omega \mathbf{M}_1 \partial\Omega \cdot \mathbf{C_E} \cdot \mathbf{C_A^T} \cdot \mathbf{U}\\
- \int_\Omega \mathbf{K_6} \partial \Omega \cdot \mathbf{C_{EQ}} = 0 
\end{eqnarray}
\end{center}

\begin{center}
\begin{eqnarray}
\int_\Omega\mathbf{N}^T \mathbf{N}\partial\Omega \cdot  \mathbf{\dot C_A}
	+ \int_\Omega \mathbf{M}_2 \partial\Omega \cdot \mathbf{C_E} \cdot \mathbf{C_S^T} \cdot \mathbf{U} = 0
\end{eqnarray}
\end{center}

\begin{center}
\begin{eqnarray} \nonumber
\int_\Omega\mathbf{N}^T \mathbf{N}\partial\Omega \cdot  \mathbf{\dot C_{EA}}
	- \int_\Omega \mathbf{M}_1 \partial\Omega \cdot \mathbf{C_E} \cdot \mathbf{C_A^T} \cdot \mathbf{U} \\
		- \int_\Omega \mathbf{K_{-2}} \partial\Omega \cdot \mathbf{C_{E^*P}}
		+ \int_\Omega \mathbf{K_2} \partial\Omega \cdot \mathbf{C_{EA}} = 0
\end{eqnarray}
\end{center}

\begin{center}
\begin{eqnarray} \nonumber
\int_\Omega\mathbf{N}^T \mathbf{N}\partial\Omega \cdot  \mathbf{\dot C_{E^*}}
- \int_\Omega \mathbf{K_3} \partial\Omega \cdot \mathbf{C_{E^*P}} \\
+ \int_\Omega \mathbf{M}_3 \partial\Omega \cdot \mathbf{C_{E^*}} \cdot \mathbf{C_B^T} \cdot \mathbf{U} = 0
\end{eqnarray} 
\end{center}

\begin{center}
\begin{eqnarray} \nonumber
\int_\Omega\mathbf{N}^T \mathbf{N}\partial\Omega \cdot  \mathbf{\dot C_{E^*P}}
- \int_\Omega \mathbf{K_{-2}} \partial\Omega \cdot \mathbf{C_{E^*P}} \\
+ \int_\Omega \mathbf{K_3} \partial\Omega \cdot \mathbf{C_{E^*P}} = 0
\end{eqnarray}
\end{center}

\begin{center}
\begin{eqnarray}
\int_\Omega\mathbf{N}^T \mathbf{N}\partial\Omega \cdot  \mathbf{\dot C_P}
- \int_\Omega \mathbf{K_3} \partial\Omega \cdot \mathbf{C_{E^*P}} = 0
\end{eqnarray}
\end{center}

\begin{center}
\begin{eqnarray}
\int_\Omega\mathbf{N}^T \mathbf{N}\partial\Omega \cdot  \mathbf{\dot C_B}
+ \int_\Omega \mathbf{M}_3 \partial\Omega \cdot \mathbf{C_{E^*}} \cdot \mathbf{C_B^T} \cdot \mathbf{U} = 0
\end{eqnarray}
\end{center}

\begin{center}
\begin{eqnarray} \nonumber
\int_\Omega\mathbf{N}^T \mathbf{N}\partial\Omega \cdot  \mathbf{\dot C_{E^*B}}
- \int_\Omega \mathbf{M}_3 \partial\Omega \cdot \mathbf{C_{E^*}} \cdot \mathbf{C_B^T} \cdot \mathbf{U} \\ \nonumber
+ \int_\Omega \mathbf{K_5} \partial\Omega \cdot \mathbf{C_{E^*B}}
+ \int_\Omega \mathbf{K_{-5}} \partial\Omega \cdot \mathbf{C_{EQ}} \\
\int_\Omega \mathbf{K_6} \partial\Omega \cdot \mathbf{C_{EQ}}  = 0
\end{eqnarray}
\end{center}

\begin{center}
\begin{eqnarray} \nonumber
\int_\Omega\mathbf{N}^T \mathbf{N}\partial\Omega \cdot \mathbf{C_{EQ}}
- \int_\Omega \mathbf{K_5} \partial\Omega \cdot \mathbf{C_{E*B}} \\
+ \int_\Omega \mathbf{K_{-5}} \partial\Omega \cdot \mathbf{C_{EQ}}
+  \int_\Omega \mathbf{K_6} \partial\Omega \cdot \mathbf{C_{EQ}} = 0
\end{eqnarray}
\end{center}

\begin{center}
\begin{eqnarray}
\int_\Omega\mathbf{N}^T \mathbf{N}\partial\Omega \cdot  \mathbf{\dot C_Q}
- \int_\Omega \mathbf{K_6} \partial\Omega \cdot \mathbf{C_{EQ}} = 0
\end{eqnarray}
\end{center}

where $\mathbf{M}_1$, $\mathbf{M}_2$, $\mathbf{M}_3$ expresses the collision probabilities
of $E-A$, $E-S$, and $E-B$, respectively.

\subsubsection*{Ping-Pong Bi-Bi with parallel pathway}

In some cases, parallel pathways as in Ping-Pong Bi-Bi, such as for
DD-carboxypeptidases \cite{Frere:1973}, being an important reaction pattern to be discretized into FEM.

\begin{center}
	$E + P  \rightharpoonup  E D \rightleftharpoons E^* P \rightharpoonup E^* + P$ \\
\end{center}
\begin{center}
	$E^* + A \rightharpoonup E^* A \rightleftharpoons E T \rightharpoonup E + T$ \\
\end{center}
\begin{center}
	$E^* + B \rightharpoonup E^* B \rightleftharpoons E C \rightharpoonup E + C$ \\
\end{center}

the finite element formulation is presented as:

\begin{center}
\begin{eqnarray} \nonumber
\int_\Omega\mathbf{N}^T \mathbf{N} \partial \Omega \cdot   \mathbf{\dot C_E}
+ \int_\Omega \mathbf{M_1} \partial\Omega \cdot \mathbf{C_E} \cdot \mathbf{C_D^T} \cdot \mathbf{U} \\
- \int_\Omega \mathbf{K_6} \partial \Omega \cdot \mathbf{C_{ET}} 
- \int_\Omega \mathbf{K_9} \partial \Omega \cdot \mathbf{C_{EC}} = 0
\end{eqnarray}		
\end{center}

\begin{center}
\begin{eqnarray}
\int_\Omega\mathbf{N}^T \mathbf{N}\partial\Omega \cdot  \mathbf{\dot C_D}
	+ \int_\Omega \mathbf{M_1} \partial\Omega \cdot \mathbf{C_E} \cdot \mathbf{C_D^T} \cdot \mathbf{U} = 0
\end{eqnarray}
\end{center}

\begin{center}
\begin{eqnarray} \nonumber
\int_\Omega\mathbf{N}^T \mathbf{N}\partial\Omega \cdot  \mathbf{\dot C_{ED}}
- \int_\Omega \mathbf{M_1} \partial\Omega \cdot \mathbf{C_{E}} \cdot \mathbf{C_D^T} \cdot \mathbf{U} \\
- \int_\Omega \mathbf{K_{-2}} \partial\Omega \cdot \mathbf{C_{E^*P}}
+ \int_\Omega \mathbf{K_{2}} \partial\Omega \cdot \mathbf{C_{ED}} = 0
\end{eqnarray}
\end{center}

\begin{center}
\begin{eqnarray} \nonumber
\int_\Omega\mathbf{N}^T \mathbf{N}\partial\Omega \cdot  \mathbf{\dot C_{E^*P}}
- \int_\Omega \mathbf{K_2} \partial\Omega \cdot \mathbf{C_{ED}} \\
+ \int_\Omega \mathbf{K_{-2}} \partial\Omega \cdot \mathbf{C_{E^*P}}
+ \int_\Omega \mathbf{K_3} \partial\Omega \cdot \mathbf{C_{E^*P}} = 0
\end{eqnarray} 
\end{center}

\begin{center}
\begin{eqnarray} \nonumber
\int_\Omega\mathbf{N}^T \mathbf{N}\partial\Omega \cdot  \mathbf{\dot C_{E^*}}
- \int_\Omega \mathbf{K_3} \partial\Omega \cdot \mathbf{C_{E^*P}} \\
+ \int_\Omega \mathbf{M}_2 \partial\Omega \cdot \mathbf{C_{E^*}} \cdot \mathbf{C_A^T} \cdot \mathbf{U}
+ \int_\Omega \mathbf{M}_3 \partial\Omega \cdot \mathbf{C_{E^*}} \cdot \mathbf{C_B^T} \cdot \mathbf{U} = 0
\end{eqnarray}
\end{center}

\begin{center}
\begin{eqnarray}
\int_\Omega\mathbf{N}^T \mathbf{N}\partial\Omega \cdot \mathbf{\dot C_P}
- \int_\Omega \mathbf{K_3} \partial\Omega \cdot \mathbf{C_{E^*P}} = 0
\end{eqnarray}
\end{center}

\begin{center}
\begin{eqnarray}
\int_\Omega\mathbf{N}^T \mathbf{N}\partial\Omega \cdot  \mathbf{\dot C_A}
+ \int_\Omega \mathbf{M}_2 \partial\Omega \cdot \mathbf{C_{E^*}} \cdot \mathbf{C_A^T} \cdot \mathbf{U} = 0
\end{eqnarray}
\end{center}

\begin{center}
\begin{eqnarray} \nonumber
\int_\Omega\mathbf{N}^T \mathbf{N}\partial\Omega \cdot  \mathbf{\dot C_{E^*A}}
- \int_\Omega \mathbf{M}_2 \partial\Omega \cdot \mathbf{C_{E^*}} \cdot \mathbf{C_A^T} \\
- \int_\Omega \mathbf{K_{-5}} \partial\Omega \cdot \mathbf{C_{ET}}
+ \int_\Omega \mathbf{K_5} \partial\Omega \cdot \mathbf{C_{E^*A}} = 0
\end{eqnarray}
\end{center}

\begin{center}
\begin{eqnarray} \nonumber
\int_\Omega\mathbf{N}^T \mathbf{N}\partial\Omega \cdot  \mathbf{\dot C_{ET}}
- \int_\Omega \mathbf{K_5} \partial\Omega \cdot \mathbf{C_{E^*A}} \\
+ \int_\Omega \mathbf{K_{-5}} \partial\Omega \cdot \mathbf{C_{ET}}
+  \int_\Omega \mathbf{K_6} \partial\Omega \cdot \mathbf{C_{ET}} = 0
\end{eqnarray}
\end{center}

\begin{center}
\begin{eqnarray}
\int_\Omega\mathbf{N}^T \mathbf{N}\partial\Omega \cdot  \mathbf{\dot C_T}
- \int_\Omega \mathbf{K_6} \partial\Omega \cdot \mathbf{C_{ET}} = 0
\end{eqnarray}
\end{center}

\begin{center}
\begin{eqnarray}
\int_\Omega\mathbf{N}^T \mathbf{N}\partial\Omega \cdot  \mathbf{\dot C_B}
+ \int_\Omega \mathbf{M}_3 \partial\Omega \cdot \mathbf{C_{E^*}} \cdot \mathbf{C_B^T} \cdot \mathbf{U} = 0
\end{eqnarray}
\end{center}

\begin{center}
\begin{eqnarray} \nonumber
\int_\Omega\mathbf{N}^T \mathbf{N}\partial\Omega \cdot  \mathbf{\dot C_{E^*B}}
- \int_\Omega \mathbf{M}_3 \partial\Omega \cdot \mathbf{C_{E^*}} \cdot \mathbf{C_B^T} \cdot \mathbf{U} \\
- \int_\Omega \mathbf{K_{-8}} \partial\Omega \cdot \mathbf{C_{EC}}
+ \int_\Omega \mathbf{K_8} \partial\Omega \cdot \mathbf{C_{EB}} = 0
\end{eqnarray}
\end{center}

\begin{center}
\begin{eqnarray} \nonumber
\int_\Omega\mathbf{N}^T \mathbf{N}\partial\Omega \cdot  \mathbf{\dot C_{EC}}
- \int_\Omega \mathbf{K_8} \partial\Omega \cdot \mathbf{C_{EB}} \\
+ \int_\Omega \mathbf{K_{-8}} \partial\Omega \cdot \mathbf{C_{EC}}
- \int_\Omega \mathbf{K_9} \partial\Omega \cdot \mathbf{C_{EC}} = 0
\end{eqnarray}
\end{center}

\begin{center}
\begin{eqnarray}
\int_\Omega\mathbf{N}^T \mathbf{N}\partial\Omega  \mathbf{\cdot C_C}
- \int_\Omega \mathbf{K_9} \partial\Omega \cdot \mathbf{C_{EC}} = 0
\end{eqnarray}
\end{center}

where $\mathbf{M}_1$, $\mathbf{M}_2$, $\mathbf{M}_3$ express the collisions probabilities of $E-D$, $E-A$ and $E-B$.

\subsubsection*{Ternary-complex mechanisms}

Ternary-complex mechanism is also common in cellular processes (e.g. DNA polymerase). In this type of enzyme, two substrates
need to link to the enzyme to form a ternary complex, either in sequence or random, with the following set of reactions: 

\begin{center}
	$E + A  \rightharpoonup  EA$ \\
\end{center}
\begin{center}
	$E + B  \rightharpoonup  EB$  \\
\end{center}
\begin{center}
	$EA + B  \rightharpoonup  EAB$ \\
\end{center}
\begin{center}
	$EB + A  \rightharpoonup  EAB$ \\
\end{center}
\begin{center}
	$EAB  \rightleftharpoons  EPQ$ \\
\end{center}
\begin{center}
	$EPQ  \rightharpoonup  EP + Q$ \\
\end{center}
\begin{center}
	$EPQ  \rightharpoonup  EQ + P$ \\
\end{center}
\begin{center}
	$EP  \rightharpoonup  E + P$ \\
\end{center}
\begin{center}
	$EQ  \rightharpoonup  E + Q$ \\
\end{center}

which inside the finite element can be expressed as:

\begin{center}
\begin{eqnarray} \nonumber
\int_\Omega\mathbf{N}^T \mathbf{N} \partial \Omega \cdot   \mathbf{\dot C_E}
+ \int_\Omega \mathbf{M_1} \partial\Omega \cdot \mathbf{C_E} \cdot \mathbf{C_A^T} \cdot \mathbf{U} \\
+ \int_\Omega \mathbf{M_2} \partial\Omega \cdot \mathbf{C_E} \cdot \mathbf{C_B^T} \cdot \mathbf{U}
- \int_\Omega \mathbf{K_2} \partial \Omega \cdot \mathbf{C_{EP}} \\
- \int_\Omega \mathbf{K_4} \partial \Omega \cdot \mathbf{C_{EQ}} = 0
\end{eqnarray}	
\end{center}

\begin{center}
\begin{eqnarray} \nonumber
\int_\Omega\mathbf{N}^T \mathbf{N}\partial\Omega \cdot \mathbf{C_A}
+ \int_\Omega \mathbf{M_1} \partial\Omega \cdot \mathbf{C_E} \cdot \mathbf{C_A^T} \cdot \mathbf{U} \\
+ \int_\Omega \mathbf{K_4} \partial \Omega \cdot \mathbf{C_{EB}} = 0
\end{eqnarray}
\end{center}

\begin{center}
\begin{eqnarray} \nonumber
\int_\Omega\mathbf{N}^T \mathbf{N}\partial\Omega \cdot \mathbf{C_B}
+ \int_\Omega \mathbf{M_2} \partial\Omega \cdot \mathbf{C_E} \cdot \mathbf{C_B^T} \cdot \mathbf{U} \\
+ \int_\Omega \mathbf{K_3} \partial\Omega \cdot \mathbf{C_{EA}} = 0
\end{eqnarray}
\end{center}

\begin{center}
\begin{eqnarray} \nonumber
\int_\Omega\mathbf{N}^T \mathbf{N}\partial\Omega \cdot  \mathbf{\dot C_{EA}}
- \int_\Omega \mathbf{M_1} \partial\Omega \cdot \mathbf{C_E} \cdot \mathbf{C_A^T} \cdot \mathbf{U} \\
+ \int_\Omega \mathbf{M_3} \partial\Omega \cdot \mathbf{C_{EA}} \cdot \mathbf{C_B^T} \cdot \mathbf{U} = 0
\end{eqnarray} 
\end{center}

\begin{center}
\begin{eqnarray} \nonumber
\int_\Omega\mathbf{N}^T \mathbf{N}\partial\Omega \cdot  \mathbf{\dot C_{EB}}
- \int_\Omega \mathbf{M_2} \partial\Omega \cdot \mathbf{C_E} \cdot \mathbf{C_B^T} \cdot \mathbf{U} \\
+ \int_\Omega \mathbf{M_4} \partial\Omega \cdot \mathbf{C_{EB}} \cdot \mathbf{C_A^T} \cdot \mathbf{U} = 0
\end{eqnarray}
\end{center}

\begin{center}
\begin{eqnarray} \nonumber
\int_\Omega\mathbf{N}^T \mathbf{N}\partial\Omega \cdot  \mathbf{\dot C_{EAB}}
- \int_\Omega \mathbf{M_3} \partial\Omega \cdot \mathbf{C_{EA}} \cdot \mathbf{C_B^T} \cdot \mathbf{U} \\ \nonumber
- \int_\Omega \mathbf{M_4} \partial\Omega \cdot \mathbf{C_{EB}} \cdot \mathbf{C_A^T} \cdot \mathbf{U}
+ \int_\Omega \mathbf{K_5} \partial\Omega \cdot \mathbf{C_{EAB}} \\
- \int_\Omega \mathbf{K_{-5}} \partial\Omega \cdot \mathbf{C_{EPQ}}	= 0
\end{eqnarray}
\end{center}

\begin{center}
\begin{eqnarray} \nonumber
\int_\Omega\mathbf{N}^T \mathbf{N}\partial\Omega \cdot  \mathbf{\dot C_{EPQ}}
- \int_\Omega \mathbf{K_5} \partial\Omega \cdot \mathbf{C_{EAB}} \\ \nonumber
+ \int_\Omega \mathbf{K_{-5}} \partial\Omega \cdot \mathbf{C_{EPQ}}
+ \int_\Omega \mathbf{K_6} \partial\Omega \cdot \mathbf{C_{EPQ}} \\
+ \int_\Omega \mathbf{K_7} \partial\Omega \cdot \mathbf{C_{EPQ}} = 0
\end{eqnarray}
\end{center}

\begin{center}
\begin{eqnarray} \nonumber
\int_\Omega\mathbf{N}^T \mathbf{N}\partial\Omega \cdot  \mathbf{\dot C_{EP}}
- \int_\Omega \mathbf{K_6} \partial\Omega \cdot \mathbf{C_{EPQ}} \\
+ \int_\Omega \mathbf{K_8} \partial\Omega \cdot \mathbf{C_{EP}} = 0
\end{eqnarray}
\end{center}

\begin{center}
\begin{eqnarray} \nonumber
\int_\Omega\mathbf{N}^T \mathbf{N}\partial\Omega \cdot  \mathbf{\dot C_{EQ}}
- \int_\Omega \mathbf{K_7} \partial\Omega \cdot \mathbf{C_{EPQ}} \\
+ \int_\Omega \mathbf{K_9} \partial\Omega \cdot \mathbf{C_{EQ}} = 0
\end{eqnarray}
\end{center}

\begin{center}
\begin{eqnarray}
\int_\Omega\mathbf{N}^T \mathbf{N}\partial\Omega \cdot  \mathbf{\dot C_P}
- \int_\Omega \mathbf{K_8} \partial\Omega \cdot \mathbf{C_{EP}} = 0
\end{eqnarray}
\end{center}

\begin{center}
\begin{eqnarray}
\int_\Omega\mathbf{N}^T \mathbf{N}\partial\Omega \cdot  \mathbf{\dot C_Q}
- \int_\Omega \mathbf{K_6} \partial\Omega \cdot \mathbf{C_{EQ}} = 0
\end{eqnarray}
\end{center}

where $\mathbf{M}_1$, $\mathbf{M}_2$, $\mathbf{M}_3$, $\mathbf{M}_4$ express the colision probabilities
of $E-A$, $E-B$, $EA-B$ and $EB-A$, respectively.

\subsubsection*{Rapid-equilibrium random mechanism}

In this mechanism, the enzyme is capable to randomly link to four different substrate (A, B, D) to form the
complex EDA or EDB, producing the different molecules P, T and C \cite{Alberty:2009}, as follows:

\begin{center}
	$E + A  \rightleftharpoons  E A + D \rightleftharpoons E D A \rightleftharpoons E + P + T$ \\
\end{center}

\begin{center}
	$E + D \rightleftharpoons E D + A \rightleftharpoons E D A$ \\
\end{center}

\begin{center}
	$E D + B \rightleftharpoons E D B \rightleftharpoons E + P + C$ \\
\end{center}

\begin{center}
	$E + B \rightleftharpoons E B + D \rightleftharpoons E D B$ \\
\end{center}

which inside the finite element can be expressed as:

\begin{center}
\begin{eqnarray} \nonumber
\int_\Omega\mathbf{N}^T \mathbf{N} \partial \Omega \cdot  \dot{\mathbf{C_A}}
- \int_\Omega \mathbf{K_{-1}} \partial \Omega \cdot \mathbf{C_{EA}} \\ \nonumber
+ \int_\Omega \mathbf{M_1} \partial\Omega \cdot \mathbf{C_E} \cdot \mathbf{C_A^T} \cdot \mathbf{U}
- \int_\Omega \mathbf{K_4} \partial \Omega \cdot \mathbf{C_{EDA}} \\
+ \int_\Omega \mathbf{M_4} \partial\Omega \cdot \mathbf{C_E} \cdot \mathbf{C_D^T} \cdot \mathbf{U} = 0
\end{eqnarray}		
\end{center}

\begin{center}
\begin{eqnarray} \nonumber
\int_\Omega\mathbf{N}^T \mathbf{N}\partial\Omega \cdot \mathbf{C_E}
- \int_\Omega \mathbf{K_1} \partial \Omega \cdot \mathbf{C_{EA}} \\ \nonumber
+ \int_\Omega \mathbf{M_1} \partial\Omega \cdot \mathbf{C_E} \cdot \mathbf{C_A^T} \cdot \mathbf{U}
- \int_\Omega \mathbf{K_3} \partial \Omega \cdot \mathbf{C_{EDA}} \\ \nonumber
+ \int_\Omega \mathbf{O_{-3}} \partial\Omega \cdot \mathbf{C_E} \cdot \mathbf{C_P^T} \cdot \mathbf{C_T}
- \int_\Omega \mathbf{K_4} \partial \Omega \cdot \mathbf{C_{ED}} \\ \nonumber
+ \int_\Omega \mathbf{M_4} \partial\Omega \cdot \mathbf{C_E} \cdot \mathbf{C_D^T} \cdot \mathbf{U}
- \int_\Omega \mathbf{K_7} \partial \Omega \cdot \mathbf{C_{EDB}} \\ \nonumber
+ \int_\Omega \mathbf{O_{-7}} \partial\Omega \cdot \mathbf{C_E} \cdot \mathbf{C_P^T} \cdot \mathbf{C_T}
- \int_\Omega \mathbf{K_8} \partial \Omega \cdot \mathbf{C_{EB}} \\
+ \int_\Omega \mathbf{M_8} \partial\Omega \cdot \mathbf{C_E} \cdot \mathbf{C_B^T}\cdot \mathbf{U}  = 0
\end{eqnarray}
\end{center}

\begin{center}
\begin{eqnarray} \nonumber
\int_\Omega\mathbf{N}^T \mathbf{N}\partial\Omega \cdot \dot{\mathbf{C_B}}
- \int_\Omega \mathbf{K_8} \partial \Omega \cdot \mathbf{C_{EB}} \\ \nonumber
+ \int_\Omega \mathbf{M_8} \partial\Omega \cdot \mathbf{C_E} \cdot \mathbf{C_B^T} \cdot \mathbf{U}
- \int_\Omega \mathbf{K_6} \partial\Omega \cdot \mathbf{C_{EDB}} \\
+ \int_\Omega \mathbf{M_6} \partial\Omega \cdot \mathbf{C_{ED}} \cdot \mathbf{C_B^T} \cdot \mathbf{U} = 0
\end{eqnarray}
\end{center}

\begin{center}
\begin{eqnarray} \nonumber
\int_\Omega\mathbf{N}^T \mathbf{N}\partial\Omega \cdot \dot{\mathbf{C_D}}
- \int_\Omega \mathbf{K_4} \partial\Omega \cdot \mathbf{C_{ED}} \\ \nonumber
+ \int_\Omega \mathbf{M_4} \partial\Omega \cdot \mathbf{C_E} \cdot \mathbf{C_D^T} \cdot \mathbf{U}
- \int_\Omega \mathbf{K_9} \partial\Omega \cdot \mathbf{C_{EDB}} \\ 
+ \int_\Omega \mathbf{M_9} \partial\Omega \cdot \mathbf{C_{EB}} \cdot \mathbf{C_D^T} \cdot \mathbf{U} = 0
\end{eqnarray} 
\end{center}

\begin{center}
\begin{eqnarray} \nonumber
\int_\Omega\mathbf{N}^T \mathbf{N}\partial\Omega \cdot \dot{\mathbf{C_{EA}}}
- \int_\Omega \mathbf{M_1} \partial\Omega \cdot \mathbf{C_E} \cdot \mathbf{C_A^T} \cdot \mathbf{U} \\ \nonumber
+ \int_\Omega \mathbf{K_{-1}} \partial\Omega \cdot \mathbf{C_{EA}}
- \int_\Omega \mathbf{K_{-2}} \partial\Omega \cdot \mathbf{C_{EDA}} \\
+ \int_\Omega \mathbf{M_2} \partial\Omega \cdot \mathbf{C_{EA}} \cdot \mathbf{C_D}^T \cdot \mathbf{U} = 0
\end{eqnarray}
\end{center}

\begin{center}
\begin{eqnarray} \nonumber
\int_\Omega\mathbf{N}^T \mathbf{N}\partial\Omega \cdot \mathbf{\dot{C}_{ED}}
- \int_\Omega \mathbf{M_4} \partial\Omega \cdot \mathbf{C_E} \cdot \mathbf{C_D^T} \cdot \mathbf{U} \\  \nonumber
+ \int_\Omega \mathbf{K_{-4}} \partial\Omega \cdot \mathbf{C_{ED}} 
- \int_\Omega \mathbf{K_{-5}} \partial\Omega \cdot \mathbf{C_{EDA}}\\ \nonumber
+ \int_\Omega \mathbf{M_5} \partial\Omega \cdot \mathbf{C_{ED}} \cdot \mathbf{C_A^T} \cdot \mathbf{U}
- \int_\Omega \mathbf{K_{-6}} \partial\Omega \cdot \mathbf{C_{EDB}} \\
+ \int_\Omega \mathbf{M_6} \partial\Omega \cdot \mathbf{C_{ED}} \cdot \mathbf{C_B^T} \cdot \mathbf{U} = 0
\end{eqnarray}
\end{center}

\begin{center}
\begin{eqnarray} \nonumber
\int_\Omega\mathbf{N}^T \mathbf{N}\partial\Omega \cdot \mathbf{\dot{C}_{EB}}
- \int_\Omega \mathbf{K_{-8}} \partial\Omega \cdot \mathbf{C_{ED}} \\ \nonumber
+ \int_\Omega \mathbf{M_8} \partial\Omega \cdot \mathbf{C_E} \cdot \mathbf{C_B^T} \cdot \mathbf{U} \\ \nonumber
- \int_\Omega \mathbf{K_{-9}} \partial\Omega \cdot \mathbf{C_{EDB}} \\
+ \int_\Omega \mathbf{M_9} \partial\Omega \cdot \mathbf{C_{EB}} \cdot \mathbf{C_D^T} \cdot \mathbf{U} = 0
\end{eqnarray}
\end{center}

\begin{center}
\begin{eqnarray} \nonumber
\int_\Omega\mathbf{N}^T \mathbf{N}\partial\Omega \cdot \mathbf{\dot{C}_{EDA}}
- \int_\Omega \mathbf{M_2} \partial\Omega \cdot \mathbf{C_{EA}} \cdot \mathbf{C_D^T} \cdot \mathbf{U} \\ \nonumber
+ \int_\Omega \mathbf{K_{-2}} \partial\Omega \cdot \mathbf{C_{EDA}}
+ \int_\Omega \mathbf{O_3} \partial\Omega \cdot \mathbf{C_E} \cdot \mathbf{C_D^T} \cdot \mathbf{C_T} \\ \nonumber
+ \int_\Omega \mathbf{K_3} \partial\Omega \cdot \mathbf{C_{EDA}}
- \int_\Omega \mathbf{M_5} \partial\Omega \cdot \mathbf{C_{ED}} \cdot \mathbf{C_A^T} \cdot \mathbf{U} \\
+ \int_\Omega \mathbf{K_{-5}} \partial\Omega \cdot \mathbf{C_{EDA}} = 0
\end{eqnarray}
\end{center}

\begin{center}
\begin{eqnarray} \nonumber
\int_\Omega\mathbf{N}^T \mathbf{N}\partial\Omega \cdot \mathbf{\dot{C}_{EDB}}
+ \int_\Omega \mathbf{M_6} \partial\Omega \cdot \mathbf{C_{EB}} \cdot \mathbf{C_B^T} \cdot \mathbf{U} \\ \nonumber
+ \int_\Omega \mathbf{K_{-6}} \partial\Omega \cdot \mathbf{C_{EDB}}
- \int_\Omega \mathbf{O_{-7}} \partial\Omega \cdot \mathbf{C_E} \cdot \mathbf{C_P^T} \cdot \mathbf{C_C} \\ \nonumber
+ \int_\Omega \mathbf{K_7} \partial\Omega \cdot \mathbf{C_{EDB}}
- \int_\Omega \mathbf{M_9} \partial\Omega \cdot \mathbf{C_{EB}} \cdot \mathbf{C_D^T} \cdot \mathbf{U} \\
+  \int_\Omega \mathbf{K_{-9}} \partial\Omega \cdot \mathbf{C_{EDB}} = 0
\end{eqnarray}
\end{center}

\begin{center}
\begin{eqnarray} \nonumber
\int_\Omega\mathbf{N}^T \mathbf{N}\partial\Omega \cdot \mathbf{\dot{C}_P}
- \int_\Omega \mathbf{K_3} \partial\Omega \cdot \mathbf{C_{EDA}} \\ \nonumber 
+ \int_\Omega \mathbf{O_{-3}} \partial\Omega \cdot \mathbf{C_E} \cdot \mathbf{C_P^T} \cdot \mathbf{C_T} 
- \int_\Omega \mathbf{K_7} \partial\Omega \cdot \mathbf{C_{EDB}} \\
- \int_\Omega \mathbf{O_{-7}} \partial\Omega \cdot \mathbf{C_E} \cdot \mathbf{C_P^T} \cdot \mathbf{C_C} = 0
\end{eqnarray}
\end{center}

\begin{center}
\begin{eqnarray} \nonumber
\int_\Omega\mathbf{N}^T \mathbf{N}\partial\Omega \cdot \mathbf{\dot{C}_T}
- \int_\Omega \mathbf{K_3} \partial\Omega \cdot \mathbf{C_{EDA}} \\
+ \int_\Omega \mathbf{O_{-3}} \partial\Omega \cdot \mathbf{C_E} \cdot \mathbf{C_P^T} \cdot \mathbf{C_T} = 0
\end{eqnarray}
\end{center}

\begin{center}
\begin{eqnarray} \nonumber
\int_\Omega\mathbf{N}^T \mathbf{N}\partial\Omega \cdot \mathbf{\dot{C}_C}
- \int_\Omega \mathbf{K_7} \partial\Omega \cdot \mathbf{C_{EDB}} \\
+ \int_\Omega \mathbf{O_{-7}} \partial\Omega \cdot \mathbf{C_E} \cdot \mathbf{C_P^T} \cdot \mathbf{C_C} = 0
\end{eqnarray}
\end{center}

where $M$ expresses the frequency of reaction of metabolisms inside the finite element. 

\subsection*{Chemical networks}

\begin{figure*}  
        \begin{center}
           \vspace{1cm}
	   \includegraphics[scale=0.45] {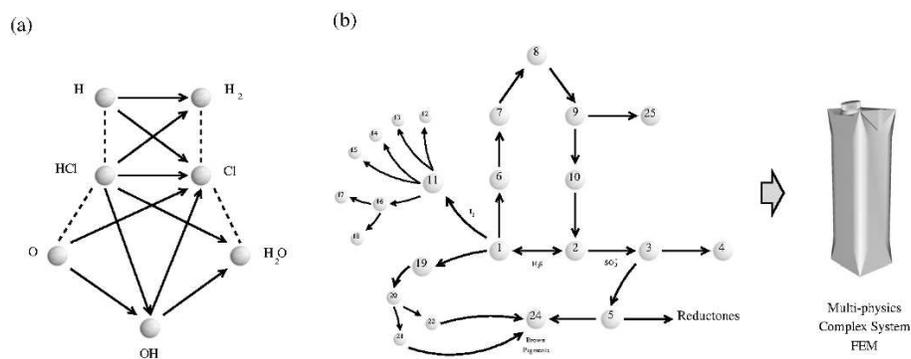}
	   \caption{Chemical system reaction network: (a) mechanistic network representation of the set of
	    chemical reactions in section 6; and (b) pseudo-mechanistic reaction network of ascorbic acid
	    degradation in foods (adapted from \cite{BauernfeindandPinkert:1970,Tannenbaum:1985})}
	  \label{figure:chemicalnetworks}
        \end{center}
\end{figure*}

When reactions are put together to describe a chemical system, it can be formalized as a graph,
where reactions are links or edges and specimens are nodes (Figure \ref{figure:chemicalnetworks}). 
Take for example the following chemical set of reactions:

\begin{center}
	$H + HCl \rightarrow H_2 + Cl$ \\
	$HCl + O \rightarrow Cl + OH$ \\
	$HCl + OH \rightarrow Cl + H_2O$
\end{center}

that can be represented by the graph in Figure \ref{figure:chemicalnetworks} (a). In this network, all reactions involve 
a second order reaction kinetics mechanism, following the FEM discretization presented in section \ref{section:2ndorder}.
If no spacial variation is considered, the differential equation for the presented reaction network is as follows:

\begin{equation}
    \mathbf{\dot{C}}  + \mathbf{S} \cdot \mathbf{C} = 0
\end{equation}

where $\mathbf{\dot{C}}$ is the reaction rate and $\mathbf{S}$ the stoichiometry matrix derived from both stoichiometry
and reaction graph. For this reaction network, the following system of equations is obtained:

\begin{tiny}
\begin{equation} \label{equation:finitedifference}
    \left[
    \begin{array}{c}
        \dot C_H \\
	\dot C_{HCL} \\
	\dot C_{H_2} \\
	\dot C_{Cl} \\
	\dot C_{O} \\
	\dot C_{OH} \\
	\dot C_{H_2O} \\
    \end{array}
    \right] + 
    \left[
    \begin{array}{ccc}
         1 & 0 & 0 \\
	 1 & 1 & 1 \\
	 -1 & 0 & 0 \\
	 -1 & -1 & -1 \\
	 0 & 1 & 0 \\
	 0 & -1 & -1 \\
	 0 & 0 & -1 \\
    \end{array}
    \right]  \cdot    \left[
    \begin{array}{c}
        k_1C_HC_{HCL} \\
	k_{2}C_{O}C_{HCL} \\
	k_3C_{OH}C_{HCL} 
    \end{array}
    \right] = 0
\end{equation}
\end{tiny}

which must be solved by optimization methods. The reaction network can also
be represented by an incidence matrix (speciemens relationships) to be used for network topology characterization 
\cite{AlbertandBarabasi:2002,DorogovtsevandMendes:2003,BarabasiandBonabeau:2003,Barabasi:2007,Barabasi:2009}. 

Once the chemical system is assumed to be in steady-state, ($\mathbf{\dot{C}}=0$):

\begin{equation}
 \mathbf{S}\mathbf{V} = 0
\end{equation}

where $\mathbf{S}$ is the stoichiometric matrix and $V$ the specimens flux vector ($mol.s-1$). The
same problem can be derived for the mass-balance of each specimen:

\vspace*{0.3cm}
\begin{tiny}
\begin{equation}
    \left[
    \begin{array}{ccccccc}
         1 & 1 & -1 & -1 & 0 & 0 & 0 \\
	 0 & 1 & 0 & -1 & 1 & -1 & 0\\
	 0 & 1 & 0 & -1 & 0 & -1 & -1
    \end{array}
    \right]  \cdot    \left[
    \begin{array}{c}
        v_H \\
	v_{HCl} \\
	v_{H_2} \\
	v_{Cl} \\
	v_{O} \\
	v_{OH} \\
	v_{H_2O} 
    \end{array}
    \right] = 0
\end{equation}
\end{tiny}

\begin{equation}
 \mathbf{R}\mathbf{V} = 0
\end{equation}

where it can be shown that $diag(k_iC_jC_k)\cdot S$ is equivalent to the 2nd term of eq \ref{equation:finitedifference}
being therefore an equivalent way of presenting reaction networks. If one considers the concentrations formulation, the
reaction network dynamical system across the physical domain is given by:

\begin{equation}
 \int_\Omega \mathbf{N}^t\mathbf{N} d\Omega \mathbf{\dot{C}} + \mathbf{S} \otimes \int_\Omega \mathbf{M} d\Omega 
\mathbf{C}_i\mathbf{C}_j^t \mathbf{U} = 0 
\end{equation}

or for a 1st order reaction kinetics:

\begin{equation}
 \int_\Omega \mathbf{N}^t\mathbf{N} d\Omega \mathbf{\dot{C}} + \mathbf{S} \otimes \int_\Omega \mathbf{K} d\Omega 
\mathbf{C}_i = 0 
\end{equation}

or by joining different mechamisms:

\begin{equation}
 \int_\Omega \mathbf{N}^T\mathbf{N} d\Omega \mathbf{\dot{C}} + \mathbf{S}_1 \otimes \int_\Omega \mathbf{M} d\Omega 
\mathbf{C}_i\mathbf{C}_j^T \mathbf{U} +  \mathbf{S}_2 \otimes \int_\Omega \mathbf{K} d\Omega 
\mathbf{C}_i = 0 
\end{equation}

Where $\mathbf{S}_1$ and $\mathbf{S}_2$ handle all the stoichiometric relationships between specimens.

In chemical systems, network reconstruction is harder to cure when compared with biochemical data.
Information is still scattered throughout publications and less efforts have been put into reconstructing
chemical systems, such as in atmospheric science and foods. For example, Figure \ref{figure:chemicalnetworks} (b)
presents part of known ascorbic acid (AA) degradation pathways \cite{BauernfeindandPinkert:1970,Tannenbaum:1985}.
The full understanding of the AA degradation has major impact on both nutrition and quality of foods,
but it still lacks the major mechanistic steps and thermodynamics. The same is valid for many important aging
and degradation mechanism which involve oxidation \cite{Martinsetal:2008c}.
The reconstruction of this network implies the existence of high-throughput analytical chemistry dedicated
facilities and bioinformatics, so that complex systems approaches can be applied to this
research area \cite{Martinsetal:2009b}.

As there is incomplete information, network simulation has to rely on flux analysis and measurements of flux rates
instead of concentrations, kinetic rates, catalysis and Arrhenius activation energies. Considering that fluxes
inside a triangular finite element is given by the shape function:

\begin{equation}
 v(\Omega) = N_iv_i + N_jv_j + N_kv_k
\end{equation}

where, $v_i$, $v_j$ and $v_k$ are the specimen flux at nodal positions $i$, $j$ and $k$; and the variational
problem is resumed to:

\begin{eqnarray} \nonumber
 V(\Omega) = \frac{1}{2} \int_\Omega \mathbf{S}v \cdot v(\Omega) d\Omega = \\  \nonumber
= \frac{1}{2} \int_\Omega \mathbf{R} v^2 d\Omega = \\
=  \frac{1}{2} \mathbf{R} \int_\Omega (N_iv_i + N_jv_j + N_kv_k)^2 d\Omega
\end{eqnarray}

that once minimised for the node $i$, helds:

\begin{equation} 
 \frac{\delta V}{\delta v_i} = \mathbf{R} \int_\Omega (N_i^2v_i + N_iN_jv_j + N_iN_kv_k) d\Omega
\end{equation}

and performing for all nodal positions and chemical specimens, is possible to conclude the final
matrix format:

\begin{equation} \label{eq:chemiomics}
 \mathbf{R} \otimes \int_\Omega \textbf{N}^T\textbf{N} d\Omega \mathbf{V} = 0
\end{equation}

Where all stoichiometric relationships inside the FE space are respected, because:

\begin{small}
\begin{equation}
    \mathbf{A} = \mathbf{N}^T\mathbf{N} = 
    \left[
    \begin{array}{ccc}
         N_i^2 & N_iN_j & N_iN_k \\
	 N_iN_j & N_j^2 & N_jN_k \\
	 N_iN_k & N_jN_k & N_k^2 
    \end{array}
    \right]  
\end{equation}
\end{small}

and therefore, eq \ref{eq:chemiomics} in network of Figure \ref{figure:chemicalnetworks} (a) is expanded to:

\begin{small}
\begin{equation}
    \int_\Omega
    \left[
    \begin{array}{ccccccc}
         \mathbf{A} & \mathbf{A} & -\mathbf{A} & -\mathbf{A} & \mathbf{0} & \mathbf{0} & \mathbf{0} \\
	 \mathbf{0} & \mathbf{A} & \mathbf{0} & -\mathbf{A} & \mathbf{A} & -\mathbf{A} & \mathbf{0} \\
	 \mathbf{0} & \mathbf{A} & \mathbf{0} & -\mathbf{A} & \mathbf{0} & -\mathbf{A} & -\mathbf{A}
    \end{array}
    \right]  d\Omega \cdot  \mathbf{V} = 0
\end{equation}
\end{small}

where $\mathbf{0}$ is a zero squared matrix, and $\mathbf{V}$ expands into a colum vector (21,1):

\begin{equation} 
  \left[V_{H_i} V_{H_j} V_{H_k} \cdots  V_{H_2O_i} V_{H_2O_j} V_{H_2O_k}\right] 
\end{equation}

Where all fluxes can be computed for any region of space.

The same problem can be discretized using the stoichiometric matrix in eq \ref{equation:finitedifference}, where in complex chemical systems can be assembled from a knowledge base
database table (Figure \ref{figure:chemicalnetworks}), where reactions, specimens, stoichiometric factors,
presence of catalysts, flux code and activation energies are cataloged, to obtain
a linear system $\mathbf{S}\mathbf{V}=0$, where in this example, $V= \left[ v_{r1}~v_{r2}~v_{r3} \right]$,
and $v_{r1}=k_1C_{H}C_{HCl}$, $v_{r2}=k_2C_{O}C_{HCl}$ and $v_{r1}=k_3C_{OH}C_{HCl}$, respectively.

It can shown that inside any finite element, the set of equations became:

\begin{equation}
 \mathbf{S} \otimes \int_\Omega \textbf{N}^T\textbf{N} d\Omega \mathbf{V}_e = 0
\end{equation}

where for a triangular finite element,

\begin{equation}
  \mathbf{V}_e = \left[ v_{r1_i}~v_{r1_j}~v_{r1_k}~\cdots~v_{r3_i}~v_{r3_j}~v_{r3_k} \right] 
\end{equation}

Being the solution for any given chemical network solve accross the physical domain as:

\begin{tiny}
\begin{equation} \label{equation:ChemicalNetworksFluxSolution}
    \int_\Omega
    \left[
    \begin{array}{ccc}
         \mathbf{A} & 0 & 0 \\
	 \mathbf{A} & \mathbf{A} & \mathbf{A} \\
	 0 & -\mathbf{A} & 0 \\
	 -\mathbf{A} & -\mathbf{A} & -\mathbf{A} \\
	 0 & \mathbf{A} & 0 \\
	 0 & \mathbf{A} & \mathbf{A} \\
	 0 & 0 & \mathbf{A} \\
    \end{array}
    \right] d\Omega  \cdot  \mathbf{V} = 0
\end{equation}
\end{tiny}

In many cases, reaction mechanisms are not fully understood and incomplete analytical chemistry may 
not allow to derive all time-course dependencies in chemical systems. For engineering purposes, empirical
pseudo-reaction steps can be assumed in incomplete reaction networks, such as for the ascorbic acid
oxidation presented in Figure \ref{figure:chemicalnetworks} (b). The same formulation is possible to be
presented to the pseudo-mechanistic network while there is not total knowledge about all 
reaction mechanisms (e.g. computational shelf-life dating \cite{Martinsetal:2008}).

\subsubsection*{Effect of temperature and catalysts}

Pure chemical systems can be considered 'auto-regulated' by thermodynamics, that is, mechanical
properties, kinetic and equilibrium constants, activation energies and presence of catalysts.
Chemical reactions dependence on temperature are generally modeled by the Arrhenius relationship:

\begin{equation}
 k = k_{ref} \times exp \left( -\frac{Ea}{R} \left[ \frac{1}{T} - \frac{1}{T_{ref}} \right] \right)
\end{equation}

where $k$ and $k_{ref}$ are the kinetic rates at temperature $T$ and $T_ref$ (K), respectively; $Ea$
the Arrhenius activation energy ($J.mol-1,K-1$). The effect of catalysts can be reflected in the decrease
of $Ea$, allowing the same reactions to occur at faster rates at lower temperatures.

In order to reflect the effect of both temperature and catalysts, a weight matrix is possible to be deduced, as the
fraction of the kinetic rate of a reaction step by it's reference kinetic rate:

\begin{equation}
 W_i = \frac{k_i}{k_{ref}}
\end{equation}

demonstrating that under steady state the integration is given by
$\mathbf{S} \cdot diag(\mathbf{W}) \cdot \mathbf{V} = 0$, with the corresponding finite element formulation:

\begin{equation}
 \mathbf{S} \cdot diag(\mathbf{W}) \otimes \int_\Omega \mathbf{N}^t\mathbf{N} d\Omega \cdot \mathbf{V}=0
\end{equation}

where $0~\leq~w_i~\leq~+\infty$. Furthermore, when $w_i=0$ the reaction step is deleted (e.g. deletion of
a catalyst),  $w_i \leq 1$ reactions are slower than the reference temperature, and $w_i \geq 1$ otherwise,
enabling to study chemical systems under different environmental conditions.

\subsubsection*{'\textit{In-Silico}' genome scale networks}

Modeling cellular growth had a significant impact on biotechnology in the pre-genomic era. Models with macroscopic
assumptions, also know as 'predictive microbiology'
(e.g. \cite{McKellar:1997,BaranyiandPin:1999,McKellarandKnight:2000}) are still used due to their simplicity
of assumptions and availability of information on kinetic data. FEM formulations were already
derived for many of these models and can be found in \cite{Martinsetal:2009b}. 

The implementation of high-throughput methodologies in molecular biology (e.g. genome sequencing, electrophoresis,
protein sequencing, mass spectroscopy, NMR), automated cellular manipulation (e.g. gene knock-out) \cite{Kingetal:2009}
and the emergence of bioinformatics, provide that gene functions, protein specificity and partial metabolic networks are
available in several species (e.g. ecoli, yeast and human) in databases
such as, BioCyc \cite{BioCyc:2012}, SGD \cite{SGD:2012}, KEGG \cite{KEGG:2012} , Reactome \cite{Reactome:2012},
UniProt \cite{UniProt:2012}. With the increasing datasets, the development and
update of holistic 'in-silico' genome-scale network draft models (GSM) has became possible to be automated
\cite{DeJonghetal:2007,Henryetal:2010} for further validation by human experts to provide 'in-silico' model organisms
(Figure \ref{figure:genomescalemodel-fem}).

There are three main types of 'in-silico' GSM models: i) interaction network models; ii) steady-state stoichiometric networks;
and iii) dynamical models (e.g. ECELL \cite{Takashietal:2003}). The latest are yet less used because of the lack of reliable
'in-vivo' kinetic data, and therefore, interaction and steady state models are dominant in bioinformatics and systems
biology analysis. Genome scale models can be further classified into non-compartmentalized and compartmentalized models
(e.g. IND750 \cite{Duarteetal:2004}, IMM940 \cite{Moetal:2009}). The second class, accounts for metabolic networks contained in the different organelles and transport
reactions between organelles, cytoplasm and extracellular space. Substantially complete models are
available for \textit{ecoli} (1260 genes, 2077 reactions, 690 of transport, 1039 metabolites), 
\textit{s. cerevisiae} (e.g.IND750, 750 genes, 648 metabolites, 1149 reactions, 297 of transport) and
many other organisms in the BIGG database \cite{Schellenbergeretal:2010}.

\subsubsection{Flux-Balance Analysis}

Considering the example network inside an organism presented in 
Figure \ref{figure:genomescalemodel-fem}, at any given position of space inside a 
finite element domain, the concentration of metabolites of the 'in-silico' organism
can be given by:

\begin{equation}
  \frac{dC}{dt} - Sv + \mu C = 0
\end{equation}

where $S$ is the stoichiometric matrix, $v$ the metabolite flux (mol/s), $\mu$ the growth rate ($s^{-1}$).
In most conditions, as kinetic constants are not available 'in-vivo', these models use the flux
instead of the traditional kinetic constants.  However, $SkC=Sv$, where $v=kC$ at a 
given time or space position. Moreover, network studies assume pseudo steady-state conditions at a given time,
that is, fluxes considered stable under short time periods, when compared to population growth and concentration
of metabolites. The problem resumes to:

\begin{equation}
\mathbf{S}\mathbf{V} = 0 
\end{equation}

which for the network model is:

\begin{tiny}
\begin{equation}
    \left[
    \begin{array}{cccccccc}
        -1 & 0 & 0 & 0 & 0 & -1 & 0 & 0 \\
	1 & -1 & 0 & 0 & 1 & 0 & 0 & 0 \\
	0 & 1 & -1 & -1 & 0 & 0 & 0 & 0 \\
	0 & 0 & 1 & 0 & 0 & 0 & -1 & 0 \\
	0 & 0 & 0 & 1 & -1 & 0 & 0 & -1 \\
    \end{array}
    \right]     \left[
    \begin{array}{c}
        v_1 \\
	v_2 \\
	v_3 \\
	v_4 \\
	v_5 \\
	b_1 \\
	b_2 \\
	b_3
    \end{array}
    \right] =      \left[
    \begin{array}{c}
        0 \\
	0 \\
	0 \\
	0 \\
	0 \\
	0 \\
	0 \\
	0
    \end{array}
    \right]
\end{equation}
\end{tiny}

Taking into consideration a consistent spacial gradient of the flux $v_a$ at any
position, the solution is given by the minimization of the variational:

\begin{equation}
 V(\Omega) = \frac{1}{2} \int_\Omega \mathbf{S}\mathbf{V}^2 d\Omega
\end{equation}

and therefore, for a given metabolite $i$, the spacial solution is given by:

\begin{equation}
 \mathbf{S} \otimes \int_\Omega \mathbf{N}^T\mathbf{N} d\Omega \mathbf{V_e} = 0
\end{equation}

where all network reactions are taken into account inside the
finite element space by using the Kronecker product with the stoichiometric
matrix. Note that $V_e$ is a column vector that spawns
all vertices's fluxes, such as, for a triangular finite element
$\mathbf{V_e}=\left[ v_{1,i}~v_{1,j}~v_{1,k}~\cdots~b_{4,i}~b_{4,j}~b_{4,k} \right]$.

This simple formulation allows to perform FBA in conjunction with multi-physics FEM or
CFD simulation in any biotechnological processes. In this sense, a state-of-the-art
genome scale model analysis can be performed now with spatio-temporal resolution
and in the complex scenario that modelers want to set-up for simulation, by integrating
FEM solutions with systems biology to provide a genome scale diagnostic at any point of
the FE mesh, such as the functionalities presented in \cite{Schellenbergeretal:2011}.

State-of-the-art GSM were designed to assist molecular biology research, assuming chemostat conditions, and
not for bioprocess or complex systems simulations. Today's GSM cannot cope with:
i) complex enzymatic mechanics; ii) kinetics and temperature effect; iii) dynamical states; iv) concentrations
of metabolites; v) temporal and spacial resolution; vi) multi-physics phenomena are not taken into account
(e.g. heat transfer, diffusion, fluid flow) and v) pathways are always assumed to be optimal, where control or
thermodynamic restrictions are implemented by flux constrains.

\begin{figure*}  
        \begin{center}
	   \includegraphics[scale=0.70] {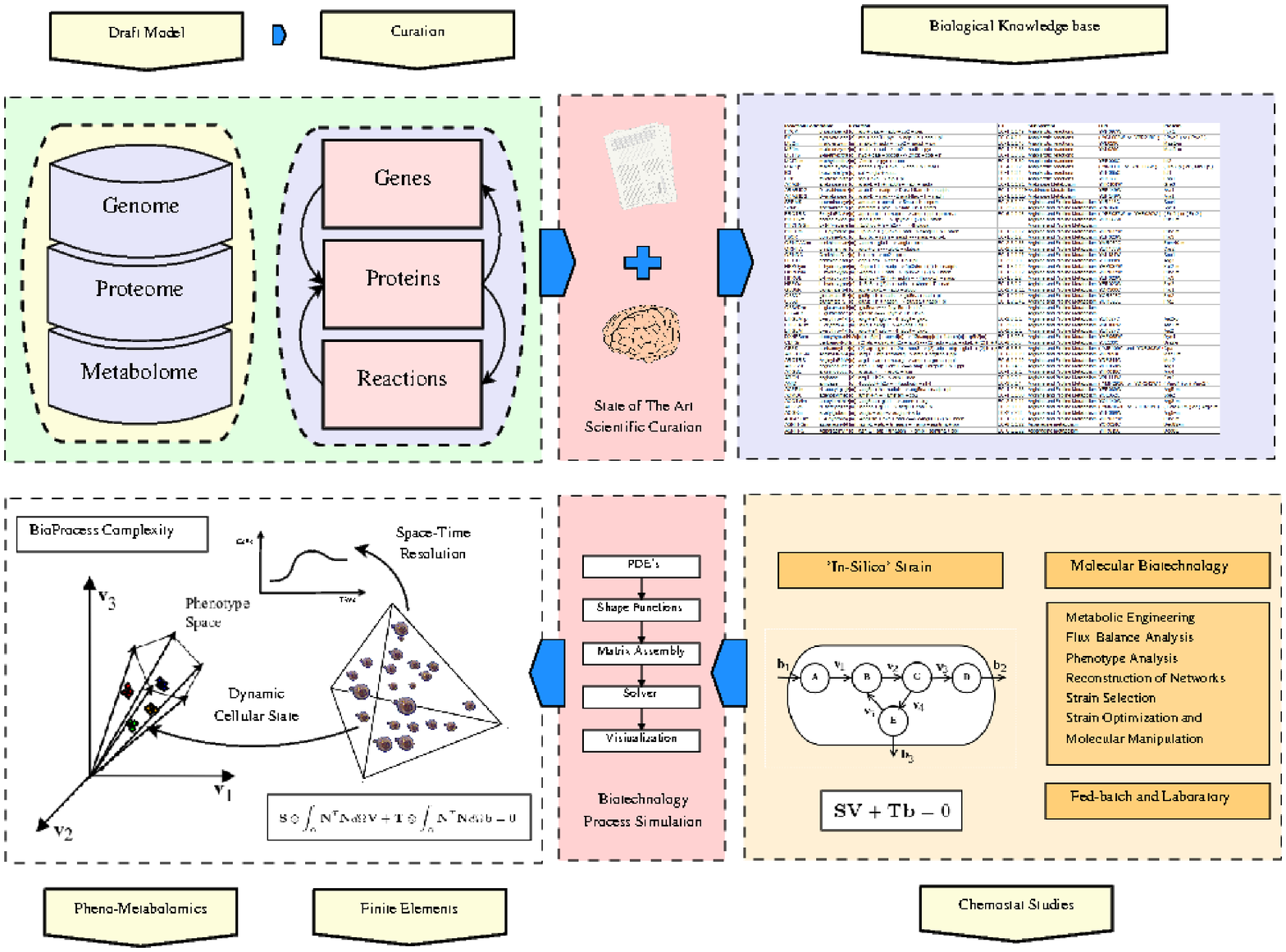}
	   \caption{Main steps for the implementation of GSM in finite elements:
	    i) development of draft models and human curation recurring to bibliography and experimental data;
	    ii) development of the knowledge base for the 'in-silico' organism implementation of the stoichiometry and
	    transport matrices, control mechanisms and flux constrains;
	    iii) assembling and solving FEM matrices for time-space resolution;
	    iv) solving and analyzing results both in the phenotype space and physical domain imaging.}
	  \label{figure:genomescalemodel}
        \end{center}
\end{figure*}

GSM provide today many applications in biotechnology, such as: i) flux balance analysis (FBA) for strain
optimization; ii) network topological analysis, reliability, viability, structural homology;
iii) derivation of phenotype spaces for the exploration of biodiversity and biotechnological potential
(Figure \ref{figure:genomescalemodel}). As these models do not hold a particular solution, both null space,
convex analysis and optimization methods are applied to explore the solution space in chemostat conditions
(e.g. MOMA, ROOM, genetic algorithms) \cite{Schellenbergeretal:2011}. Furthermore, as solutions may converge into different
regions of the phenotype, being necessary to develop new space basis, such as, the development of
elementary flux analysis \cite{GayenandVenkatesh:2006,Kimetal:2008,Beurton-Aimaretal:2011} and
extreme pathways \cite{Papinetal:2002,Priceetal:2003,Familietal:2005}.

The integration of GSM with FEM allows to overcome many of the previously mentioned barriers, allowing to perform
genome-scale analysis of cells in the context of spatio-temporal conditions in a multi-physics environment
\cite{Martinsetal:2009b}. Figure \ref{figure:genomescalemodel} exemplifies the integration. GSM are a set
of incidence matrices, computationally derived from databases and cured with publications and expert
analysis, relating genes to enzymes, enzymes and reactions, and, reactions to metabolites which given the
stoichiometric relationships can be expressed as internal and boundary fluxes of metabolites.

When deriving the GSM inside the finite element, the 'in-silico' organism becomes dependent on the external
conditions of nutrients, temperature, fluid flow, as well as, being affected by neighboring cells in any part of the physical and
time domains. FEM considers that GSM is continuously discretized across the physical domain;
at any point of the physical domain all metabolite fluxes and phenotype space is possible to be characterized,
such as, for example the coordinates inside the convex hull given by the extreme pathways
(see Figure \ref{figure:genomescalemodel}, with limitless applications in biotechnology.

\subsubsection*{Compartimented models}

In fully compartmentalized GSM models, each cellular organelle has an internal metabolic network, enzymes
and associated genes. Common metabolites among compartments are linked by transport fluxes
\cite{Duarteetal:2004,Moetal:2009}. In this reasoning, steady state equations resume to:

\begin{equation}
 \mathbf{S}\mathbf{V} + \mathbf{T}\mathbf{b} = 0 
\end{equation}

where $\mathbf{T}$ is the transport incidence matrix and $\mathbf{b}$ the boundary fluxes.
After concatenation of all organelles metabolism and transport equations, cellular
state inside a FE space is given by:

\begin{equation}
 \mathbf{S} \otimes \int_\Omega \mathbf{N}^T\mathbf{N} d\Omega \mathbf{V} +
\mathbf{T} \otimes \int_\Omega \mathbf{N}^T\mathbf{N} d\Omega \mathbf{b} = 0
\end{equation}

Being by this equation characterizes 'in-silico' compartimentalized organisms at any region of the finite
element space $\Omega$.

\subsubsection*{Pheno-metabolomics}

Pheno-metabolomics plays a major role in post-genomic biotechnology. The exploration of the phenotype and metabolic
capacities of organisms with the aid of both high-throughput methods in conjunction with genome
scale models and complex systems simulation tools lies at the heart of pheno-metabolomics bioinformatics.
Our research center has an important biodiversity yeast biobank, with especial emphasis on
\textit{Saccharomyces cerevisiae} isolates, and has been working in the characterization of
\textit{S. cerevisiae} over the last decade of yeast from different ecological contexts and geographical
origins for their phenotype potential \cite{SchullerandCasal:2007,Franco-Duarteetal:2009}.



The pheno-metabolome of species is highly diversified, but most particular solutions
of GSM have been restricted to the validation of simple, controlled experimental conditions \cite{Duarteetal:2004}
which do not reflect the complexity of real-world bioprocess and natural conditions where yeasts
evolved, lacking the design of new tools to both detect and derive new mechanisms as well as to cope
with the dynamical complexity of cells. The use of GSM has been restricted to the
assessment of the phenotype space derived from the stoichiometric matrix, being necessary to develop new
approaches to fully explore the biodiversity of biobanks, evolution and adaptation mechanisms, as
well as, the discovery of unknown mechanism by integration of GSM with both high-throughput signal
processing, statistical computing, process analytical technology and computational simulation in order
to be possible do derive the most correct definition of the phenotype space.

\begin{figure*} [hf]
        \begin{center}
           \vspace{1cm}
	   \includegraphics[scale=0.3] {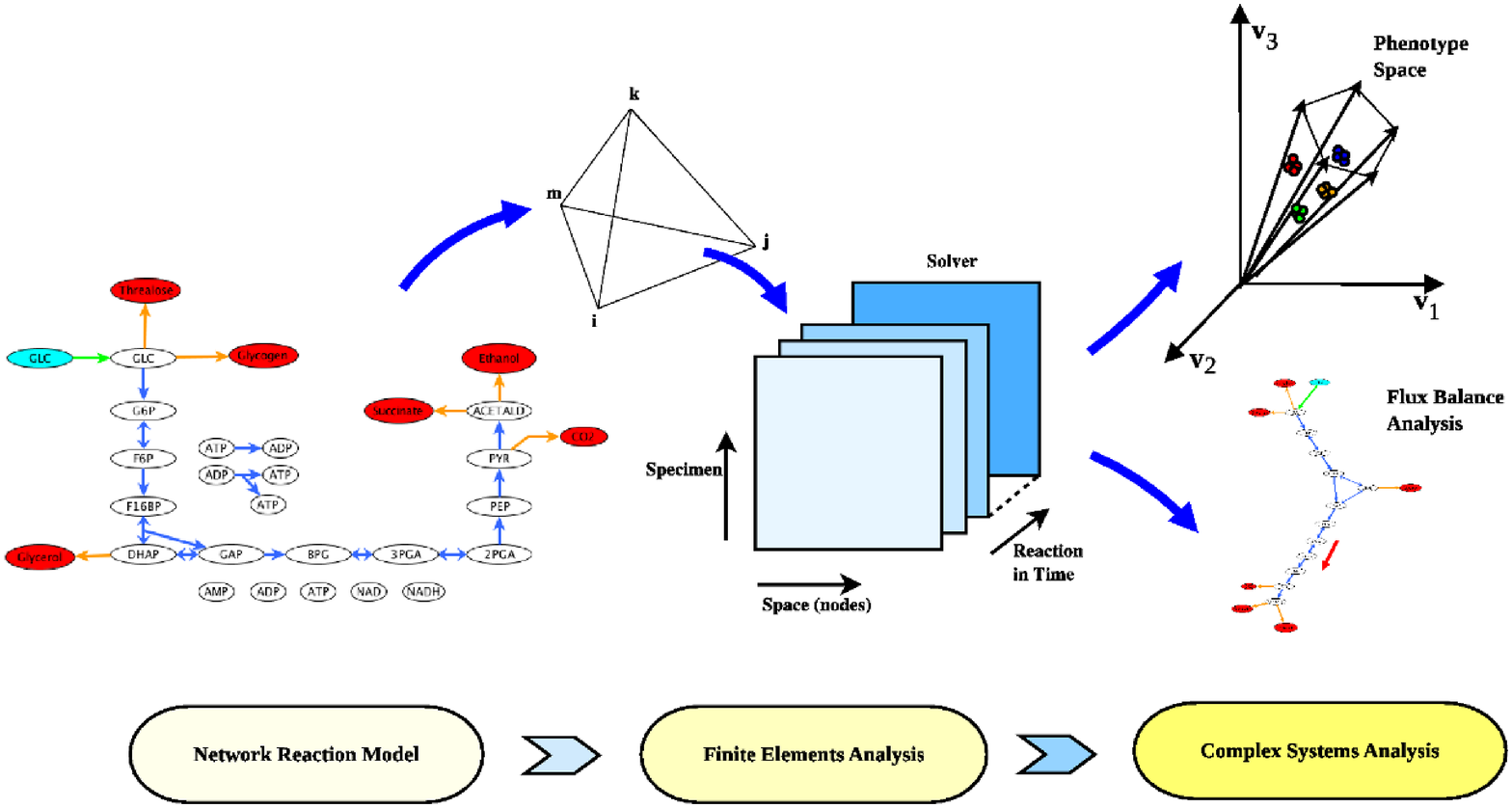}
	   \caption{Key integration steps of reaction network models and finite elements: spacio-temporal
discretization on FE space, time-course computation and results analysis in the phenotype space and fluxes,
given different GSM configurations and environmental conditions.}
           \label{figure:IntegrativeComplexSystems}
        \end{center}
\end{figure*}

One of the first approaches to define the phenotype of species was proposed as a non-negative linear combination
of all relationships present in the stoichiometric matrix, holding all non-negative possible solutions
of $\mathbf{S}\mathbf{v}=0$, when all fluxes $v_i \geq 0$ \cite{Schillingtal:2000}. Such geometry
is defined by the non-negative combination of a new vector basis, forming a convex hull defined by
extreme rays (or pathways):

\begin{equation}
 \mathbf{p} = \{v: v=\sum_i^n w_iP_i, ~ w_i \geq 0 \}
\end{equation}

where $\mathbf{p}$ is the convex space (see Figure \ref{figure:IntegrativeComplexSystems}) delimited by
the extreme pathways $P_i$ and $w_i$ the coordinates projected into each $P_i$. Note that $\mathbf{P}$ is not
an orthogonal basis, and only delimits the solution space of $\mathbf{S}\mathbf{v}=0$, being the
vectors $P_i$ presented in Figure \ref{figure:IntegrativeComplexSystems} in the natural basis of
$v_i$, which is not a practical visualization method once most GSM are hyper-dimensional.
$\mathbf{P}$ can be obtained by the methodology presented in \cite{Schillingtal:2000}, and hold
important properties for the interpretation of the phenotype space: i) primary metabolism linked
to boundary fluxes; ii) futile cycles with link to boundary fluxes; and iii) internal cycles.

Inside a FE, the convex hull coordinates $w$ of any point are possible to be described by the
element shape function (or in any other basis):

\begin{equation}
 w_e = N_iw_i + N_jw_j + N_kw_k
\end{equation}
 
allowing to apply finite element analysis (FEA) techniques do diagnose space differentiation in 
phenotype and metabolic state on the extreme pathways vector basis $P_i$.

\subsubsection*{Spacio-temporal analysis}

Spacio-temporal analysis is perhaps one of the major advantages of joining FEM and GSM, becaming possible
to analyze how the metabolic state evolves throughout space-time, as well as,
to access how different phenotypes respond to different environment conditions. Previous sections already
presented how to include fluxes ($v_i$) and pheno-metabolome coordinates ($w_i$) on a finite element domain.
Such allows to analyze emergent patterns in cell communities and perform systems biology analysis \cite{Schellenbergeretal:2011}
at each region of space the cause of phenotype differences. Such tool will become more and more important,
as cellular morphology may became manageable inside bioreactors \cite{Castroetal:2009,Silvaetal:2009}. 

For instance, the use of the FEM allows to derive space vector gradients of both fluxes and phenotypes:

\begin{equation}
  \frac{du_e}{dt} = \frac{dN_i}{d(x,y,z)} u_i + \frac{dN_j}{d(x,y,z)} u_j + \frac{dN_k}{d(x,y,z)} u_k
\end{equation}

where $u_e$ is the gradient property to be analysed across the physical domain.

FEA may be used to further explore the phenotype dynamics, where for example the space derivate allows to
determine geometrical changes in phenotypes across the FE domain (e.g. change rate ($\frac{dw_e}{dt}$) and
acceleration ($\frac{dw_e}{dt}$)):

\begin{eqnarray}
    \frac{du_e}{dt} = N_i\frac{du_i}{dt} + N_j\frac{du_j}{dt} + N_k\frac{du_k}{dt} \\
    \frac{d^2u_e}{dt^2} = N_i\frac{d^2u_i}{dt^2} + N_j\frac{d^2u_j^2}{dt} + N_k\frac{d^2u_k}{dt^2}
\end{eqnarray}

Allowing to explore dynamically the molecular biology of different phenotypes, such as, the determination
of the most important pathways and cellular functions at different stages, understand enzyme efficiency and
metabolic rates, regulation mechanisms and transcription rates in different contexts of cellular
growth, as well as, understanding accelerations in phenotype changes or metabolic states as adaptations
to changes in the environment. Figure \ref{figure:genomescalemodel} resumes the use of the phenotype
coordinates with FEM.

As the solution of GSM equations is in many cases stochastic \cite{Schellenbergeretal:2011}, it is also important
to be able to visualize the statistics of predictions in the FE domain. For example, is possible to derive both
expected phenotype $\hat{w}_e$ and corresponding variance $\sigma^2 (w)$ on a surface:

\begin{equation}
     \hat{w}_e = \frac {\int_s \mathbf{N}\mathbf{w} ds}{s}
\end{equation}

\begin{equation}
     \sigma^2 (w) = \frac {\int_s (\mathbf{N}\mathbf{w}-\hat{w}_e )^2 ds}{s}
\end{equation}

where $s$ is the finite element surface area ($m^2$) \cite{Martinsetal:2009b}.

As this new approach may provide many possible solutions in the phenotype space, and therefore inverse FEM methods
coupled with real-time and high-throughput experimental methods in molecular biology will be necessary to
fine tune the numerical results of FEA analysis. Table \ref{table:FemInPheno-Metabolomics} presents 
analogies between FEM-GSM and biological implications. Moreover, as dynamical results can be complex in terms of
interpretation, pattern recognition recurring to compressed space coordinates may be more appropriate
than direct visualization of fluxes and phenotype coordinates.

The integration of FEM with reaction networks and genome scale networks will play an important role in the
simulation and diagnostic of complex biological systems in the near future. Systems biology and systems chemistry
lacked the possibility of integrating systems knowledge with multi-physics and multi-scale physics with
4D discretization that may enable in the future the computational assessment of phenotype tests, such as
diagnostic the metabolic states under different growth media, emergence effects of gene deletion and stress
factors, as well as bioengineering issues such as, reactor temperature, must composition and bioreactor
design. This kind of tools will also open new possibilities in deriving and exploring the phenotype space
for effective exploration of biobanks, providing critical informations for the decision of strain selection or
improvement for a given biotechnological process. This manuscript is an introduction to the endless
possibilities that are open for both study of complexity by FEM and network models and use of this methodology
for the exploration of phenotypes, diagnosis, modeling, simulation and control of complex bioprocesses.

\section*{Acknowledgments}
This work was finantially supported by the Funda\c{c}\~ao para a Ci\^encia e Tecnologia,
projects OpenMicrobio (PTDC\-/BIO\-/69310\-/2006) - A framework for the simulation of cellular communities
during bioprocess engineering, Phenomet (PTDC\-/AGR-ALI\-/103392\-/2008 FEDER\-/COMPETE).
The funders had no role in study design, data collection and analysis, decision to publish, or preparation of
the manuscript.

\bibliography{artdb,wwwdb,bookdb,chpdb,thesis,procd,mandb}

\clearpage
\listoffigures
\newpage

\clearpage
\newpage

\clearpage
\newpage
\begin{figure*} 
        \begin{center}
           \vspace{1cm}
	   \includegraphics[scale=0.80] {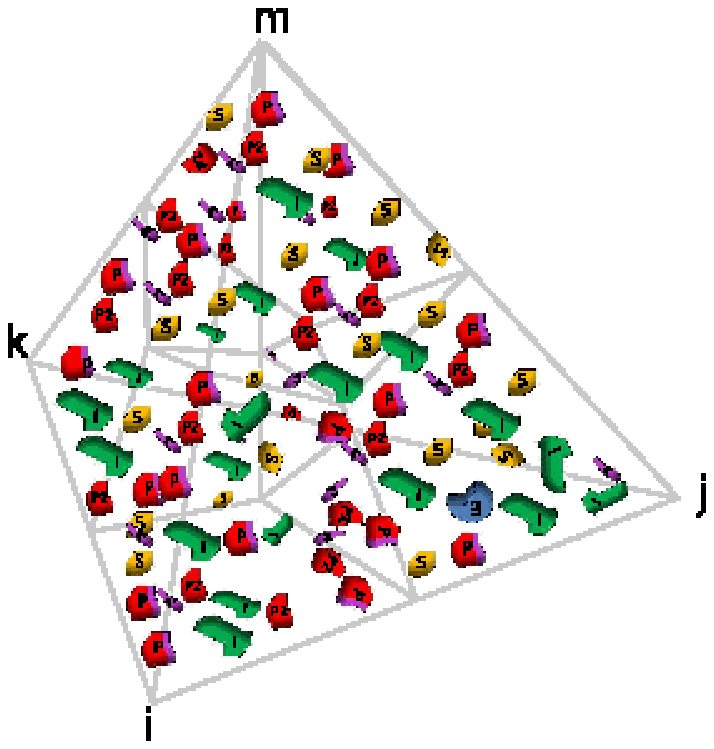}
	   \includegraphics[scale=0.80] {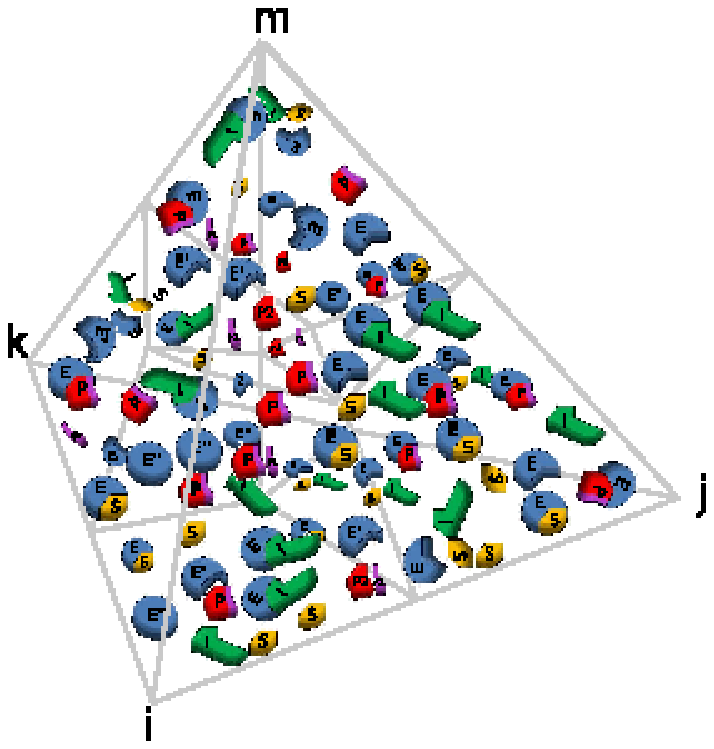} \\
	   \includegraphics[scale=0.80] {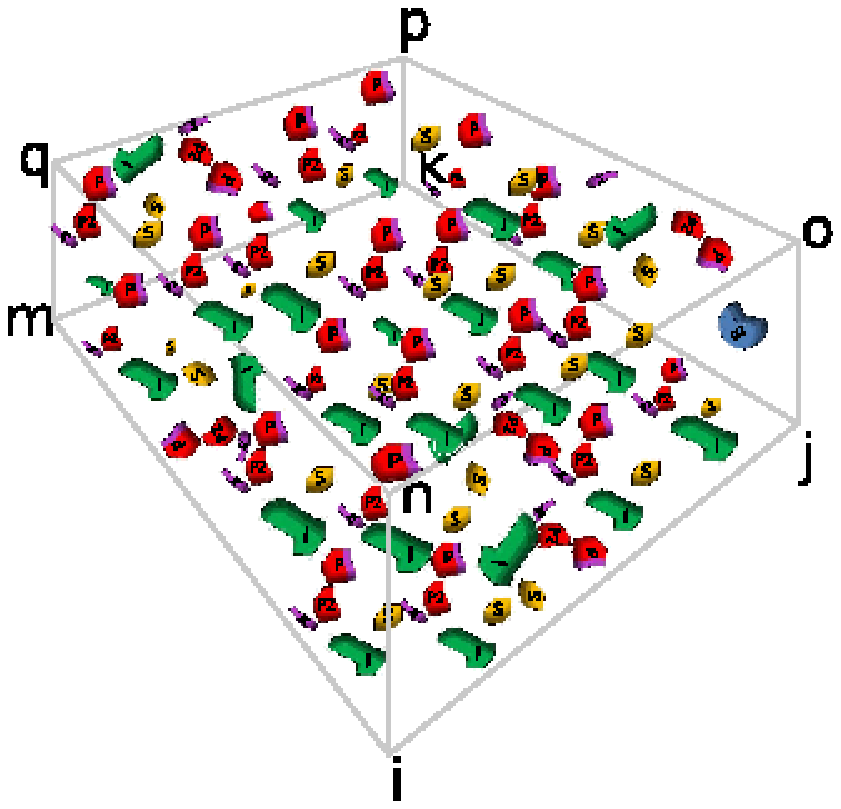}
	   \includegraphics[scale=0.80] {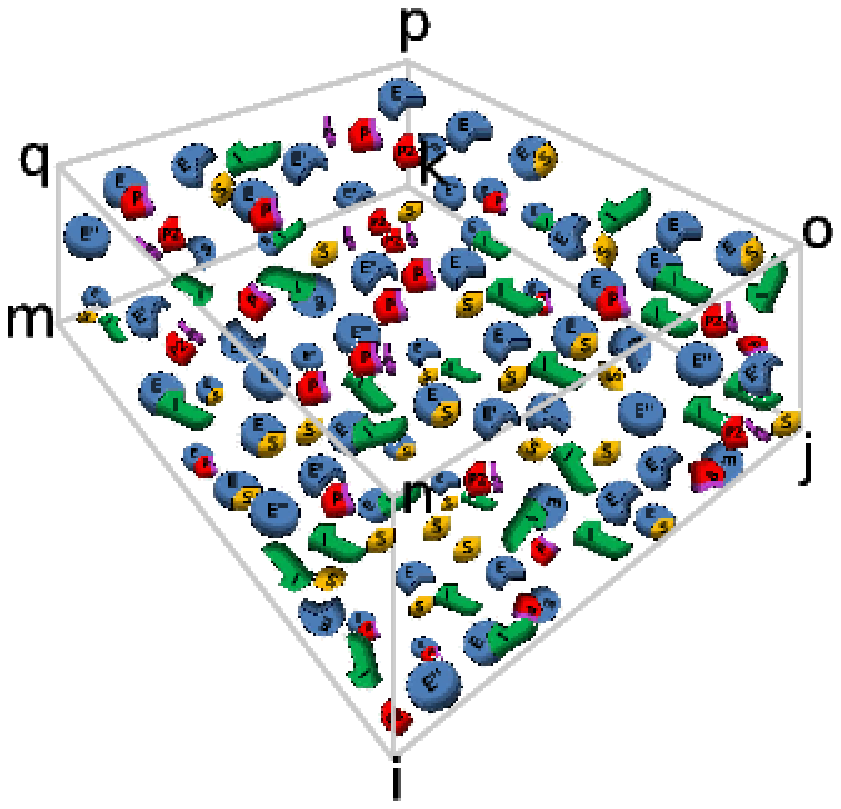}
	   \caption{Example illustrating the FE concept with competitive inhibition: reaction rates are a continuous
probabilistic function inside the FE space as function of concentration, given by approximation by the
element shape function, providing a piecewise solution in the physical domain.}
           \label{figure:randomreaction}
        \end{center}
\end{figure*}

\clearpage
\newpage

\clearpage
\newpage

\clearpage
\newpage
\begin{figure*}  
        \begin{center}
           \vspace{1cm}
	   \includegraphics[scale=0.85] {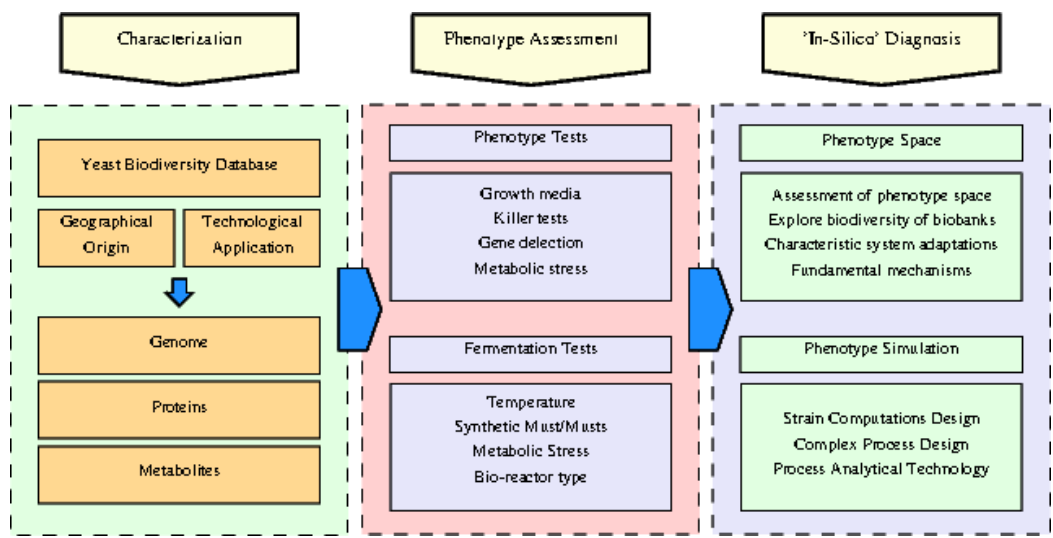}
	   \caption{FEM applications for yeast pheno-metabolome exploration in biotechnology.}
	  \label{figure:FutureApplications-GSM-FEM}
        \end{center}
\end{figure*}

\clearpage
\newpage
\listoftables

\newpage
\begin{table} [ht]
\caption{Finite elements interpolation polynomials and shape functions} \label{table:FEShapeFunctions}
\begin{center}
\begin{tabular}{lll}
Element  & Interpolation Polynomial & Shape Function  \\ \hline
Linear Beam & $u(x)=\gamma_1+\gamma_2x$ & $u = N_1u_1 + N_2u_2$ \\
Linear Triangle & $u(x,y)=\gamma_1+\gamma_2x+\gamma_3y$ & $u = N_1u_1 + N_2u_2 + N_3u_3$ \\
Linear Quadrilateral & $u(x,y)=\gamma_1+\gamma_2x+\gamma_3y + \gamma_4xy$ & $u = N_1u_1 + N_2u_2 + N_3u_3 + N_4u_4$ \\
Linear Tetrahedron & $u(x,y,z)=\gamma_1+\gamma_2x+\gamma_3y + \gamma_4z$ & $u = N_1u_1 + N_2u_2 + N_3u_3 + N_4u_4$ \\ 
Linear Cube & $u(x,y,z)=\gamma_1+\gamma_2x+\gamma_3y + \gamma_4z$ & $u = N_1u_1 + \ldots + N_8u_8$ \\
& $+ \gamma_5xy+\gamma_6xz+\gamma_7yz + \gamma_8xyz$ & \\ \hline
Quadratic Beam & $u(x)=\gamma_1+\gamma_2x+\gamma_3x^2$ & $u = N_1u_1 + N_2u_2$ \\
Quadratic Triangle & $u(x,y)=\gamma_1+\gamma_2x+\gamma_3y$ & $u = N_1u_1 + \ldots + N_6u_6$ \\
& $+ \gamma_4x^2 + \gamma_3y^2 + \gamma_3xy$ &  \\
Quadratic Quadrilateral & $u(x,y)=\gamma_1+\gamma_2x+\gamma_3y$ & $u = N_1u_1 + \ldots + N_8u_8$ \\
& $+ \gamma_4x^2 + \gamma_5y^2 + \gamma_6xy$ &  \\
& $+ \gamma_7x^2y + \gamma_8xy^2$ &  \\
Quadratic Tetrahedron & $u(x,y,z)=\gamma_1+\gamma_2x+\gamma_3y + \gamma_4z$ & $u = N_1u_1 + \ldots + N_7u_7$ \\
& $+ \gamma_5x^2 + \gamma_6y^2 + \gamma_7z^2$ &  \\
Quadratic cube & $u(x,y,z)=\gamma_1+\gamma_2x+\gamma_3y + \gamma_4z$ & $u = N_1u_1 + \ldots + N_{20}u_{20}$ \\
& $+ \gamma_5xy + \gamma_6xz + \gamma_7yz + \gamma_8xyz$ &  \\
& $+ \gamma_9x^2 + \gamma_{10}y^2 + \gamma_{11}z^2 + \gamma_{12}x^2y$ &  \\
& $+ \gamma_{13}x^2z + \gamma_{14}xy^2 + \gamma_{15}y^2z + \gamma_{16}xz^2y$ &  \\
& $+ \gamma_{17}yz^2 + \gamma_{18}x^2yz + \gamma_{19}xy^2z + \gamma_{16}xyz^2$ &  \\ \hline
\end{tabular}
\end{center}
\end{table}

\begin{table} [ht]
\caption{Finite elements interpolation polynomials and shape functions} \label{table:FemInPheno-Metabolomics}
\begin{center}
\begin{tabular}{lll}
Finite Element  & $\longrightarrow$ & Biology  \\ \hline
$w_e$ & $\longrightarrow$ & Phenotype spacial distribution  \\
$\hat{w}_e,\sigma (w)$ & $\longrightarrow$ & Phenotype statistical distribution  \\
$\frac{dw_e}{dt}$ & $\longrightarrow$ & Rate of cellular differentiation  \\
$\frac{d^2w_e}{dt^2}$ & $\longrightarrow$ & Rate of cellular adaptation  \\ 
$\nabla w$ & $\longrightarrow$ & Phenotype spacial differentiation vector \\ \hline
\end{tabular}
\end{center}
\end{table}

\end{document}